\documentclass[aps,prx,superscriptaddress,longbibliography,twocolumn]{revtex4-2}

\pdfoutput=1

\usepackage{amssymb}
\usepackage{amsmath}
\usepackage{amsthm}
\usepackage{graphicx,bm}
\usepackage{epstopdf}
\usepackage{subfigure}
\usepackage{natbib}
\usepackage{epsfig}
\usepackage{amsfonts}
\usepackage{mathrsfs}
\usepackage{CJK}
\usepackage[toc,page,title,titletoc,header]{appendix}
\usepackage{array,longtable}
\usepackage{float}
\usepackage[usenames, dvipsnames, x11names]{xcolor}
\usepackage[colorlinks]{hyperref}
\usepackage[capitalise]{cleveref}
\usepackage{enumerate}
\usepackage{soul}
\usepackage{dsfont}
\hypersetup{
 colorlinks=true,
 citecolor=red,
 linkcolor=blue,
 urlcolor=cyan}
\usepackage{mathtools}
\usepackage{lipsum}

\newcommand{\tr}{\mathrm{tr}}

\newcommand{\R}{\mathbb{R}}

\renewcommand{\L}{\hat{\mathcal{L}}}

\newcommand{\dd}{\ \mathrm{d}}

\newcommand{\grad}{\nabla}

\newcommand{\Z}{\mathbb{Z}}

\newcommand{\C}{\mathbb{C}}

\newcommand{\1}{\mathds{1}}

\renewcommand{\Re}{\mathrm{Re}}
\newcommand{\hc}{\mathrm{H.c.}}

\newcommand{\E}{\mathds{E}}

\newcommand{\Heff}{\hat H_{\mathrm{eff}}}
\newcommand{\BarGamma}{\bar{\Gamma}}

\usepackage{comment}

\begin{document}

\title{Efficient simulation of non-trivial dissipative spin chains via stochastic unraveling}
\author{Andrew Pocklington}
\email{abpocklington@uchicago.edu}
\affiliation{Department of Physics, University of Chicago, 5640 South Ellis Avenue, Chicago, Illinois 60637, USA }

\affiliation{Pritzker School of Molecular Engineering, University of Chicago, Chicago, IL 60637, USA}

\author{Aashish A. Clerk}
\email{aaclerk@uchicago.edu}
\affiliation{Pritzker School of Molecular Engineering, University of Chicago, Chicago, IL 60637, USA}

\date{\today}

\begin{abstract}
We present a new technique for efficiently simulating (in polynomial time) a class of one-dimensional dissipative spin chains that,  when mapped to fermions, have quadratic Hamiltonians, with the only nonlinearity coming from Jordan-Wigner strings appearing in the jump operators, despite the fact that these models cannot be mapped to quadratic fermionic master equations.  We show that many such Lindblad master equations admit an exact stochastic unraveling with individual trajectories evolving as Gaussian fermionic states, even though the full master equation describes a system inequivalent to free fermions.  This allows one to calculate arbitrary observables efficiently without sign problems, and with bounded sampling complexity. We utilize this new technique to study three paradigmatic dissipative effects: the melting of anti-ferromagnetic order in the presence of local loss, many-body subradiant phenomenon in systems with correlated loss, and non-equilibrium steady states of a 1D dissipative transverse-field Ising model. Beyond simply providing a powerful numerical technique, our method can also be used to gain both qualitative and quantitative insights into the role of interactions in these models.
\end{abstract}


\maketitle 


\section{Introduction}
\label{sec:intro}

One dimensional spin chains are paradigmatic models in theoretical physics, informing our understanding of a plethora of phenomena, including transport \cite{Bertini2021,Landi2022}, quantum phase transitions \cite{Sachdev1999,Giamarchi2003}, and entanglement dynamics \cite{Calabrese2005,DeChiara2006,Hastings2007}.
Extending these systems to an open systems context with driving and dissipation gives rise to a wealth of new effects, and is also crucial for understanding the many-body dynamics of many NISQ-era quantum simulation platforms, where coupling to the environment can never be neglected \cite{Preskill2018}.
For some 1D spin models, the Hamiltonian can be solved exactly via a Jordan-Wigner mapping to a system of free fermions \cite{Jordan1928}. However, adding even innocuous-looking forms of dissipation (e.g.~local spin relaxation) can break the integrability of the system, as the linear spin operators appearing in dissipators pick up non-trivial Jordan-Wigner strings. Hence, outside of a handful of models that have been solved exactly (see e.g.~\cite{Prosen2011,Prosen2013,Yao2024}), describing the effects of even simple dissipative processes can require an exponential computational overhead.

\begin{figure}
    \centering
    \includegraphics[width=\linewidth]{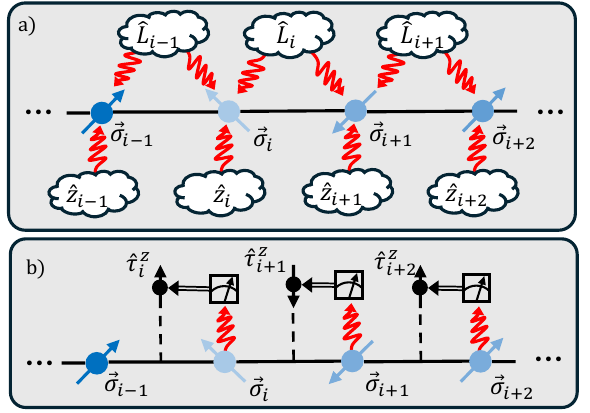}
    \caption{a) Schematic of the class of models simulable using our techniques: a 1D spin chain with linear dissipation that is local or correlated between sites (jump operators $\hat z_i$ or $\hat{L}_i$, respectively), and a Hamiltonian with anisotropic $XY$ nearest-neighbor exchange and possibly non-uniform local $Z$ fields.  Upon a Jordan-Wigner transformation, due to non-trivial string operators in the dissipators, the system is mapped to a fermionic model inequivalent to free fermions. b)  We can interpret the action of the string operators in the dissipation as a measurement and feedback process, where following each jump, a local $\Z_2$ gauge field undergoes a classical bit flip.
    }
    \label{fig:diagram2}
\end{figure}

In this work, we will present a new technique to efficiently simulate a class of Markovian dissipative spin chains which are non-trivial, in the sense that they cannot be mapped to quadratic fermionic master equations. We focus primarily on 1D systems where the Hamiltonian is equivalent to free-fermions, but the presence of Jordan-Wigner strings in the dissipative terms makes the system no longer free; however, we will relax this constraint later to show the technique can also work on a broader class of systems, even including 2D models.

Our method is based on an exact stochastic unraveling of the master equation, chosen so that each individual trajectory is a time-dependent fermionic Gaussian state.  This provides an efficient means for computing the dynamics of observables, even though the average state is not a Gaussian state and the full dynamics isn't free fermion. Our method possesses many advantageous features:  it has provable convergence results, has a computational complexity that scales only polynomially in the number of spins, and is free of any sign problems because it samples from only physical trajectories \cite{Kamar2025}. Because our technique is not limited to low-entanglement states (indeed, Gaussian states can have arbitrarily high amounts of entanglement), we believe our method can compliment existing simulation techniques that rely on area laws to achieve polynomial scaling.

The magic underlying our approach is that for a class of interesting master equations, we can find a stochastic unraveling that directly maps to a kind of fermionic measurement-plus-feedback dynamics, see \cref{fig:diagram2}.  In this mapping, non-trivial string operators are now interpreted as feedback unitaries that follow a quantum jump, with each unitary  having the simple form of a Gaussian gauge transformation.  Using this correspondence, we show that our technique is equivalent to a noisy $\Z_2$ gauge theory where the gauge fields are \textit{classically} dynamical (i.e.~they always have definite values), see \cref{fig:diagram2}. This has the benefit of both making the dynamics efficient to simulate, as well as providing direct physical insights into the effect that string operators have on the dynamics that would not be as easy to elucidate from other simulation techniques. This also allows us to also expand the technique beyond simple 1D spin chains to more complicated $\Z_2$ gauge theories, such as the Kitaev Honeycomb in 2D \cite{Kitaev2006}. 

The remainder of this paper is organized as follows: in \cref{sec:prev_work}, we introduce open spin chains and briefly review previously known results on simulation complexity. Then, in \cref{sec:new_method}, we present the most general type of spin system which we are able to simulate efficiently, and present two different algorithms loosely based on either interpreting the dynamics as being in the Heisenberg or Schr\"odinger picture. In \cref{sec:numerics}, we give a few initial examples of models we can solve. We study the melting of anti-ferromagnetic order in a 1D XX spin chain with loss, we look at the dynamics and steady state of a dissipative Ising model, and we study many-body subradiant decay in a 1D chain of emitters. Finally, in \cref{sec:boson_fermion}, we present a generalization of our technique to generic, open fermion models where the interaction is not a Jordan-Wigner string or inspired by an underlying spin model.

\section{Open Spin Chains: Previous Work}
\label{sec:prev_work}

There has been significant previous work studying open quantum spin chains, both numerically and exactly, which we will give an extremely brief outline of here. The goal will be to set the context for our solution technique, and outline how the new class of models we can solve fits into the space of previously known results.

We will be interested exclusively in Gorini-Kossakowski-Sudarshan-Lindblad (GKSL) form master equations, which represent the most generic possible completely positive, trace preserving, Markovian quantum master equations \cite{Lindblad1976,Gorini1976}. These have the form 
\begin{align}
    \partial_t \hat \rho &= -i[\hat H, \hat \rho] + \sum_i \hat L_i \hat \rho \hat L_i^\dagger - \frac{1}{2} \{ \hat L_i^\dagger \hat L_i, \hat \rho \}, \label{eqn:lind}
\end{align}
where $\hat H$ is the Hamiltonian of the system, and the collection of operators $ \hat L_i $ are ``jump'' operators describing the interaction with a Markovian environment. We will study primarily 1D spin chains by mapping them to fermions via the Jordan-Wigner (JW) transformation \cite{Jordan1928}. If we have a set of spin-1/2 operators $\hat \sigma_i^{x,y,z}$ indexed by a lattice position $i = 1, \dots, N$, then we can define a set of canonical fermion annihilation operators
\begin{align}
    \hat \sigma_i^- &= \left( \prod_{j = 1}^{i - 1} (-1)^{\hat c_j^\dagger \hat c_j} \right) \hat c_i \equiv \hat \Pi_{i - 1} \hat c_i,
\end{align}
where $\hat \sigma_i^- = (\hat \sigma_i^x + i \hat \sigma_i^y)/2$. These fermions obey the canonical anti-commutation relations $\{ \hat c_i, \hat c_j \} = \{ \hat c_i^\dagger, \hat c_j^\dagger \} = 0$ and $\{ \hat c_i, \hat c_j^\dagger \} = \delta_{ij}$. The product of fermion parity operators $\hat \Pi_i$ are called JW strings.

By using such a mapping, one can directly relate the simulability of a spin problem to whether or not it maps onto a simulable system of fermions, which has been well studied in the literature \cite{Terhal2001,Valiant2001,Shtanko2021}. Considering first the coherent (Hamiltonian) dynamics), the set of nearest neighbor spin interactions that map to quadratic Hamiltonians of the fermion operators are of the form:
\begin{align}
    \hat H &= \sum_i \frac{\Delta_i}{2} \hat \sigma_i^z - J_i \hat \sigma_i^+ \hat \sigma_{i +1}^- - \tilde J_i \hat \sigma_i^+ \hat \sigma_{i +1}^+ + \hc \nonumber \\
    &= \sum_i \Delta_i \hat c_i^\dagger \hat c_i + J_i \hat c_i^\dagger \hat c_{i +1} + \tilde J_i \hat c_i^\dagger \hat c_{i +1}^\dagger + \hc, \label{eqn:JWHam}
\end{align}
where $\Delta_i \in \R$ and $ J_i, \tilde J_i \in \C$. The dynamics of any Gaussian state (see \cref{app:gaussiantogaussian} for definitions of Gaussian states/operators) under a quadratic Hamiltonian remains Gaussian, and therefore time evolving the state can be reduced to time evolving the covariance matrix of the Gaussian state.  This reduces the problem of evolving an exponentially large state vector or density matrix, to the dramatically simpler problem of evolving a polynomially-large covariance matrix. 

When adding in dissipation and again mapping to fermions, for a Gaussian state to remain Gaussian, we again need the Hamiltonian to be quadratic in fermions.  However, we also require that each jump operator $\hat L_i$ be linear in fermion creation/annihilation operators. One can immediately see that this will be problematic when starting from the spin problem: linear spin operators pick up a JW string under mapping to fermions, and so time evolution with linear spin operators does not preserve Gaussianity. For example, simple loss modeled by a jump operator $\hat L_i = \hat \sigma_i^-$ maps to $\hat L_i = \hat \Pi_{i-1} \hat c_i$. Moreover, as the string operators $\hat \Pi_i$ are extremely non-local, the resulting interactions in the fermionized model would appear to be very long range and hence intractable.

There are a few circumstances in which spin chains (with quadratic Hamiltonians) remain efficiently simulable in the presence of dissipation. If the jump operators are quadratic and hermitian (like the terms appearing in \cref{eqn:JWHam}, an example being dephasing noise), then the dynamics are equivalent to a classically stochastic quadratic Hamiltonian \cite{Breuer2002,Gardiner2004,wiseman2009}.Specifically, in these scenarios the state is not Gaussian, but the equation of motion for all $2r$-point correlation functions close on themselves, making them efficiently simulable \cite{Horstmann2013,Barthel2022}. When noise is sufficiently strong, it can damp out long range correlations such that the system undergoes a percolation transition and then becomes simulable \cite{Aharonov2000,Trivedi2022}. However, outside of these particular cases, there is no efficient and exact method for simulating what happens when the jump operators have strings.

\section{Stochastic Simulation Method}
\label{sec:new_method}

\subsection{Stochastic Unraveling of the Master Equation}

In this section, we will outline a new simulation technique which extends the set of efficiently simulable spin models. Specifically, we will demonstrate how to simulate a class of models where the non-Hermitian Hamiltonian encoded in Eq.~\eqref{eqn:lind} maps to an operator quadratic in JW fermions, but where fermionic interactions arise through string operators associated with the quantum jump terms. We stress that the techniques presented below are numerically \textit{exact} - given sufficient averaging they recover unbiased expectation values of the time continuous master equation. This is true regardless of the relative strength of the noise and whether or not it is classical. Moreover, the technique is entirely agnostic to the total amount of entanglement in the system. 

We study systems with Hamiltonians of the form \cref{eqn:JWHam} and jump operators which can be written as sums of 2-local linear spin operators:
\begin{align}
    \hat L_i &= l_{i,1} \hat \sigma_i^- + l_{i,2} \hat \sigma_i^+ + l_{i,3} \hat \sigma_{i + 1}^- + l_{i,4} \hat \sigma_{i + 1}^+ \nonumber \\
    &=  \hat \Pi_i (-1)^{\hat c_i^\dagger \hat c_i} \left( l_{i,1}  \hat c_i + l_{i,2} \hat c_i^\dagger \right) + \hat \Pi_i \left( l_{i,3} \hat c_{i + 1} + l_{i,4} \hat c_{i + 1}^\dagger \right) \nonumber \\
    &=
    \hat \Pi_i \left( l_{i,1} \hat c_i - l_{i,2} \hat c_i^\dagger + l_{i,3} \hat c_{i + 1} + l_{i,4} \hat c_{i + 1}^\dagger \right) \equiv \hat \Pi_i \hat L'_i, \label{eqn:jump_op}
\end{align}
where $l_{i,j} \in \C$ are arbitrary possibly time-dependent coefficients. With this constraint the only non-quadratic interaction terms in the fermionic representation of the master equation arise via the overall JW string operator in the jump operators
$\hat L_i$. It is crucial that we can factor out a single string operator, which is why we are limited to 2-local jumps.

We will make this class of systems efficiently simulable by rewriting the master equation as a stochastic average. We will unravel the master equation in the ``quantum jumps'' picture. We can group together the commutator and the anti-commutator in \cref{eqn:lind} to form the non-Hermitian effective Hamiltonian $\hat H_\mathrm{eff} = \hat H - \frac{i}{2} \sum_i \hat L_i^\dagger \hat L_i$. We next introduce the stochastic equation of motion \cite{Breuer2002,Gardiner2004,wiseman2009}
\begin{align}
d|\psi \rangle &= -i\hat H_\mathrm{eff} |\psi \rangle dt + \sum_i \left( \frac{\hat L_i}{\sqrt{\langle \hat L_i^\dagger \hat L_i \rangle}} - 1 \right) |\psi \rangle d \xi_i, \label{eqn:sse}
\end{align}
where $d\xi_i \in \{0,1\}$ is a point process with mean $\overline{d \xi_i} = \langle \hat L_i^\dagger \hat L_i \rangle dt$, which implicitly depends on the state at a time $t$. We can interpret the dynamics governed by \cref{eqn:sse} as the state evolving under the non-Hermitian Hamiltonian $\Heff$, with this smooth evolution interrupted with discrete quantum jumps:  at each instant in time, with a probability $\langle \hat L_i^\dagger \hat L_i \rangle dt$, the system undergoes a discrete jump defined by the jump operator $\hat L_i$. Note that this time evolution generically does not preserve the norm of $|\psi \rangle$ because $\Heff \neq \Heff^\dagger$.

The first crucial observation that we make is that
as we have constrained our jump operators to have the form $\hat L_i = \hat \Pi_{i-1} \hat L_i'$ where $\hat \Pi_{i-1}$ is a string operator and $\hat L_i'$ is a linear fermionic operator, then $\hat L_i^\dagger \hat L_i = (\hat L_i')^\dagger \hat L_i'$.  It follows that the the effective Hamiltonian $\Heff$ is a quadratic fermionic operator. This in turn implies that all of the complexity in the dynamics of each individual stochastic trajectory arises from the finite ``jumps'': the continuous time evolution under the effective Hamiltonian is purely Gaussian.  We will show below that the situation is even more favourable:  because of the specific way string operators appear (i.e.~they are exponentials of quadratic fermion operators), they will also be easily simulable. Specifically, we will show that each jump operator maps Gaussian states to Gaussian states.  We can  thus view the jump process as a Gaussian feedback process where we apply a conditional Gaussian transformation conditioned on whether or not a quantum jump occurred, see \cref{fig:diagram2}.

We will use this observation to derive two techniques to efficiently simulate the master equation. The first is roughly based on a ``Heisenberg Picture'' interpretation, where we directly solve for the time evolution of operators instead of the state. This method provides an appealing intuitive picture of the effect of the interactions encoded in the strings. The algorithm has a computational complexity that is polynomial in the number of spins, but still turns out not to be the most efficient way to simulate the dynamics. The second is based on a ``Schr\"odinger Picture'' interpretation, where the state itself is dynamic, and is more numerically efficient than the Heisenberg picture interpretation. In the remainder of this paper we provide intuition and examples of the different types of models that can be simulated, and in \cref{app:mostgeneral} we give the most general 1D spin chain model that can be simulated with our technique. Moreover, the technique can be extended to higher dimensional systems in certain cases, see \cref{app:2d}. 

Before proceeding, we note that Ref.~\cite{Torres2014} derived a result on Lindblad dynamics that applies to a 
{\it subset} of our general class of simulable models.  If we restrict attention to models where the Hamiltonian is strictly (fermionic) number conserving, and dissipation only involves loss, then Ref.~\cite{Torres2014} provides a method for computing the spectral decomposition of the Liouvillian (using its lower-triangular, block-diagonal structure). The spectrum can be obtained from $\Heff$ alone, and this can then be used to recursively calculate eigenmodes. We stress that this result does not apply to the more general class of models we consider, where the Hamiltonian need not conserve excitation number, and where dissipation can involve particle loss and addition.  Even in the more restricted loss-only setting, our approach remains extremely valuable, as using the results of Ref.~\cite{Torres2014} to compute the dynamics of observables still results in an exponentially hard problem (see \cref{app:classicalDifficulty} for more details).  

\subsection{Heisenberg Picture}
\label{subsec:Heis}

On any given stochastic trajectory, the system evolves under $\Heff$, punctuated by quantum jumps.
Consider time-evolution from $t = 0$ until $t = t_f$
starting from a pure state $| \psi \rangle$
, during which a specific trajectory undergoes $m$ quantum jumps.  We index these jumps with $j = 1, \dots, m$, and use $\hat L_{i_j}$ to denote the jump operator corresponding to jump $j$. For this specific trajectory, the expectation value of an arbitrary observable $\hat{O}$ at time $t = t_f$ takes the form:
\begin{widetext}
\begin{align}
    \langle \hat O \rangle(t_f) &= \langle \psi | e^{i \Heff^\dagger t_1} \hat L_{i_1}^\dagger \dots e^{i \Heff^\dagger (t_m - t_{m-1})} \hat L_{i_m}^\dagger e^{i \Heff^\dagger (t_f - t_m)} \hat O e^{-i \Heff (t_f - t_m)} \hat L_{i_m} e^{-i \Heff (t_m - t_{m-1})} \dots  \hat L_{i_1} e^{-i \Heff t_1} | \psi \rangle/\mathcal{N}(t),
\end{align}
where $\mathcal{N}(t)$ corresponds to state normalization. 
To help elucidate the structure here, we introduce the set of operators $\hat P_j = \hat \Pi_{i_1} \hat \Pi_{i_2} \dots \hat \Pi_{i_j}$ (i.e.~the product of string operators associated with jumps $1,2,...,j$).  We define the Heisenberg picture time evolution for an operator under the non-Hermitian Hamiltonian evolution as:
\begin{align}
    \partial_t  \hat O_\mathrm{unnorm} &= i \left( \Heff^\dagger \hat O_\mathrm{unnorm} - \hat O_\mathrm{unnorm} \Heff \right),  \\
    \partial_t \mathcal{N} &= \partial_t \langle \1 \rangle = i \langle \Heff^\dagger -\Heff \rangle,
\end{align}
where $\hat O  =  \hat O_\mathrm{unnorm}/\mathcal{N}$. 

We next introduce a time-dependent non-Hermitian Hamiltonian whose form is set by the pattern of jumps occurring along a specific trajectory of interest:
\begin{align}
\Heff'(t) = \left\{
\begin{array}{cc}
\Heff & 0 < t < t_1 \\
\hat P_j \Heff \hat P_j &  t_j < t < t_{j  +1}
\end{array} \right., \label{eqn:timeDepHam}
\end{align}
with $t_{m + 1} \equiv T$.
Note crucially that as each string operator $\hat \Pi_i$ is Gaussian (i.e., an exponential of quadratic fermion operators), so are the $\hat P_j$ operators.  It follows that for any trajectory, the corresponding $\Heff'(t)$ is a quadratic fermionic operator for all times. 
 With these definitions, our expectation value takes the convenient form:
\begin{align}
    \langle \hat O \rangle(T) &= \langle \psi | (\hat L')_{i_1}^\dagger(t_1) \dots ( \hat L')_{i_m}^\dagger (t_m) (\hat P_m \hat O \hat P_m)(t_f) \hat L'_{i_m}(t_m)  \dots  \hat L'_{i_1} (t_1) | \psi \rangle / \mathcal{N}(t).
\end{align}
\end{widetext}
Recall that the $\hat{L}'_j$ are our jump operators {\it without} the overall string-operator prefactor (c.f.~\eqref{eqn:jump_op}), and note that the Heisenberg-picture operators here evolve according to the trajectory-dependent non-Hermitian Hamiltonian $\Heff'(t)$

We see that we have a string of time-evolved operators, where each 
$\hat L'_{i_j} (t_j)$ will be linear in fermionic raising and lowering operators (as it is a linear-in-fermion operator evolved under a {\it quadratic} Hamiltonian).  This is a very tractable structure as long as our initial state $|\psi\rangle$ is Gaussian:  it corresponds to computing an out-of-time-order correlator (OTOC) of a \textit{Gaussian state}, which can be done efficiently. More explicitly, as each $\hat L'_{i_j} (t_j)$ is linear in fermion operators, this OTOC can be efficiently computed by diagonalizing $\Heff$ and then computing a Pfaffian, see \cref{app:complexity} for details. This gives the expectation value for the observable $\hat O$ over a single stochastic trajectory; the observable for the master equation can be computed by averaging over the trajectories. Overall, this can be done with asymptotic computational complexity $\mathcal{O}(N^7)$ where $N$ is the number of spins in the chain, proving that observables can be efficiently simulated, see \cref{app:complexity} for more details.

Eq.~\eqref{eqn:timeDepHam} for our time-dependent conditional Hamiltonian has a straightforward interpretation: after each quantum jump, we flip some of the signs in the effective Hamiltonian based on the position of the jump, and this modified Hamiltonian then governs the evolution until the next jump.  These sign-flips directly correspond to the string operators effecting a position-dependent gauge transformation $\hat c_j \to - \hat c_j$ for all of the sites to the left of the jump.
We thus obtain a useful intuitive picture for the action of the string operators in the dissipators:  they drive a kind of time-dependent disorder that is correlated with the location of the jumps.

We can make the above picture more explicit by gauging the weak $\Z_2$ symmetry in the problem. We first introduce a set of auxiliary spins $\hat \tau_i^\alpha$ ($\alpha=x,y,z$) in each jump operator of our original spin master equation via:
\begin{align}
    \hat L_{i,G} &= \hat \tau_i^x \left( l_{i,1} \hat \sigma_i^- + l_{i,2} \hat \sigma_i^+ + l_{i,3} \hat \sigma_{i + 1}^- + l_{i,4} \hat \sigma_{i + 1}^+ \right).
\end{align}
The auxiliary spins do not appear in the Hamiltonian.  
Since $\hat \tau_i^x$ is a strongly conserved quantity, they have no impact on the dynamics. Now, when we map to fermions, we find that
\begin{align}
    \hat H_{G} &= \sum_i \Delta_i \hat c_i^\dagger \hat c_i + \hat{\tilde \tau}_i^z \left(  J_i \hat c_i^\dagger \hat c_{i +1} +   \tilde J_i \hat c_i^\dagger \hat c_{i +1}^\dagger \right) + \hc, \\
    \hat{\tilde \tau}_i^z &\equiv (i \hat \gamma_{\tau,i} \hat{\bar \gamma}_{\tau,i}), \\
    \hat L_{i,G} &= \hat \gamma_{\tau,i} \left( l_{i,1} \hat c_i - l_{i,2} \hat c_i^\dagger + l_{i,3} \hat c_{i + 1} + l_{i,4} \hat c_{i + 1}^\dagger \right),
\end{align}
where $\hat \gamma_{\tau,i}, \hat{\bar \gamma}_{\tau,i}$ are Majorana  fermions. 
The gauge factors $\hat{\tilde \tau}_i^z$ that now appear in the Hamiltonian controls the sign of the interaction on each bond \cite{Smith2017}, and are dynamical due to the dissipation.  However, upon a stochastic unraveling, the dynamics of the $\hat{\tilde \tau}_i^z$  become completely classical, because the jump terms all commute or anticommute with the local gauge degrees of freedom
\begin{align}
    [\hat L_{i,G}, \hat{\tilde \tau}_j^z] &= 0 \ \ \forall i \neq j, \\
    \{ \hat L_{i,G}, \hat{\tilde \tau}_i^z \} &= 0.
\end{align}
Hence, we can treat the gauge degrees of freedom as numbers $\hat{\tilde \tau}_i^z = \pm 1$ that are classically updated after each jump. This is just another way of rewriting \cref{eqn:timeDepHam} where instead of the Hamiltonian being time-dependent, there is instead a classically dynamical gauge field fluctuating between $1 \leftrightarrow -1$ after each quantum jump.

It is also interesting to point out that this interpretation of a conditional gauge transformation being applied by each quantum jump is an interesting type of non-reciprocal feedback identified in \cite{Wang2023}, where each jump applies a unitary gate on a different subsystem. This interpretation will prove useful in later sections to understand some of the specific models we consider, and how they differ from their free fermion counterparts. Moreover, this interpretation allows us to generalize the simulation technique to higher dimensional $\Z_2$ gauge theories, for example the Kitaev honeycomb, which we show can be simulated with noise in \cref{app:2d}.

\subsection{Schr\"odinger Picture}

While the previous Heisenberg-picture formulation provides useful intuition in terms of effective time-dependent disorder, it is not the most efficient way to do simulations.  We now present things in the Schr\"odinger picture, where only the state vector changes in time.  To achieve this, we make use of a second key observation: despite the string operator prefactor, the action of each jump operator $\hat L_i$ 
[c.f.~\cref{eqn:jump_op}] maps Gaussian states into Gaussian states. To see this, we will decompose it into two factors, and show that each is Gaussianity-preserving.  First, we can consider the string operator $\hat \Pi_i$. This is trivially Gaussian, as it can equivalently be written as $\hat \Pi_i = \exp(i \pi \sum_{j = 1}^i \hat c_i^\dagger \hat c_i )$, which is just a time evolution unitary generated by a quadratic Hamiltonian.  The remaining factor $\hat{L}'_i$ in our jump operator is linear in fermionic raising and lowering operators. Acting with such  linear fermionic operators also maps Gaussian states into Gaussian states (see \cref{app:gaussiantogaussian} for details).

We assume throughout that our initial state is Gaussian.  Since the state $|\psi \rangle$ remains Gaussian along each individual trajectory (as defined in \cref{eqn:sse}), it suffices to only keep track of its covariance matrix. It will be useful to do this in a Majorana basis, we hence define:
\begin{align}
    \hat \gamma_{2i-1} &= \hat c_i + \hat c_i^\dagger, \\
    \hat \gamma_{2i} &= i(\hat c_i - \hat c_i^\dagger),
\end{align}
such that $\{ \hat \gamma_i, \hat \gamma_j \} = 2 \delta_{ij}$. We then define the $2N \times 2N$ covariance matrix as
\begin{align}
    \Gamma_{ij} &= \langle \hat \gamma_i \hat \gamma_j \rangle.
\end{align}

Our problem thus reduces to computing the evolution of $\Gamma_{ij}(t)$ along each trajectory, due both to $\Heff$ and the discrete quantum jumps. 
Consider first the no-jump evolution. 
If we allow $\Heff$ to be a general quadratic Hamiltonian defined by an antisymmetric matrix $H$: 
\begin{align}
    \Heff &= \sum_{ij} H_{ij} \hat \gamma_i \hat \gamma_j,
\end{align}
then we can define $X = 4iH$ the generator of the dynamics of $\Gamma$ via the non-linear equation of motion
\begin{align}
    \partial_t \Gamma &= \Gamma X - X^* \Gamma + \frac{1}{2}\Gamma(X^* - X)\Gamma.
    \label{eq:GammaEOM}
\end{align}
When $\Heff$ is Hermitian, then $X = X^*$ and we recover the standard equations of motion for the fermionic covariance matrix. The nonlinear term encodes the fact the Hamiltonian is not Hermitian, and the state norm is changing. 

Next, we include the effects of quantum jumps.  Note that a given JW  string operator $\hat \Pi_m$ can be written as 
\begin{align}
    \hat \Pi_m &= i^m \prod_{j = 1}^m \gamma_{2 j-1} \gamma_{2j}.
\end{align}
If we write the non-string part of each jump operator as
\begin{align}
    \hat L_i' &= \sum_j l_{ij} \hat \gamma_j,
\end{align}
then following a jump involving $\hat L_i$, the state goes from $|\psi \rangle \to |\psi'\rangle =  \hat L_i |\psi \rangle/\sqrt{\langle \hat L_i^\dagger \hat L_i \rangle}$. The covariance matrix after the jump is defined as  $\Gamma_{mn}' = \langle\psi'| \hat \gamma_m \hat \gamma_n | \psi' \rangle$. We can write this covariance matrix in terms of the pre-jump covariance matrix $\Gamma_{mn}$:
\begin{align}
    \Gamma_{mn}' &= \frac{\sum_{kl} l^*_{ik} l_{il} \langle \hat \gamma_k \hat \Pi_i \hat \gamma_m \hat \gamma_n \hat \Pi_i \hat \gamma_l \rangle}{\sum_{kl} l^*_{ik} l_{il} \langle \hat \gamma_k \hat \gamma_l \rangle},
\end{align}
where the expectations values are with respect to the state pre-jump. This can be simplified greatly by noticing that as $\hat \Pi_i$ is a string of Majorana's, it either commutes or anticommutes with $\hat \gamma_m$ and $\hat \gamma_n$, depending on the sign of $m - 2i$. We can collect all of these phases together and define
\begin{align}
    \eta^i_{m} = \left\{ 
    \begin{array}{cc}
       -1  & m \leq 2i \\
       +1  & m > 2i
    \end{array}
    \right.,
\end{align}
such that
\begin{align}
    \hat \Pi_i \hat \gamma_m &= \eta^i_m  \hat \gamma_m \hat \Pi_i.
\end{align}
Formally, these phases are keeping track of the conditional gauge transformations being introduced by the string operators. In \cref{subsec:Heis} we introduced gauge operators $\hat{\tilde{\tau}}_m^z$, here we can observe that $\langle \hat{\tilde{\tau}}_m^z \rangle \to \eta^i_m \langle \hat{\tilde{\tau}}_m^z \rangle$ following a quantum jump at a site $i$. Using these phase variables, we find:
\begin{align}
    \Gamma_{mn}' &= \frac{\sum_{kl} l^*_{ik} l_{il} \eta^i_m \eta^i_n \langle \hat \gamma_k  \hat \gamma_m \hat \gamma_n \hat \gamma_l \rangle}{\sum_{kl} l^*_{ik} l_{il} \langle \hat \gamma_k \hat \gamma_l \rangle} \nonumber \\
    &= \frac{\sum_{kl} l^*_{ik} l_{il} \eta^i_m \eta^i_n (\Gamma_{km}\Gamma_{nl} + \Gamma_{kl}\Gamma_{mn} - \Gamma_{kn} \Gamma_{ml})}{\sum_{kl} l^*_{ik} l_{il} \Gamma_{kl}}. \label{eqn:update}
\end{align}
Hence, this gives us a simple way to update the covariance matrix of the conditional state after a given quantum jump.
Note that the string operators enter this update rule through the phase factors $\eta^i_m \eta^i_n$.  This corresponds to the effective unitary ``feedback" operation occurring after each fermionic jump (see ~\cref{fig:diagram2}).

\cref{eq:GammaEOM,eqn:update} completely specify how to evolve the covariance matrix along a particular stochastic trajectory.
We can conveniently combine these rules into a single stochastic equation of motion for the covariance matrix:
\begin{widetext}
\begin{align}
    \dd \Gamma_{mn} &= \left( \Gamma X - X^* \Gamma + \frac{1}{2} \Gamma(X^* - X)\Gamma \right)_{mn} \dd t + \sum_i\left( \frac{\sum_{kl} l^*_{ik} l_{il} \eta^i_m \eta^i_n (\Gamma_{km}\Gamma_{nl} + \Gamma_{kl}\Gamma_{mn} - \Gamma_{kn} \Gamma_{ml})}{\sum_{kl} l^*_{ik} l_{il} \Gamma_{kl}} - \Gamma_{mn} \right) \dd \xi_i.     \label{eq:StochasticGamma}
\end{align}
\end{widetext}
As before, the $\dd \xi_i$ are independent, nonlinear point processes that depend on the current state of the system, satisfying $\dd \xi_i^2 = \dd \xi_i$ and $\overline{\dd \xi_i} = \langle \hat L_i^\dagger \hat L_i \rangle \dd t$.

In \cref{app:complexity} we provide 
an explicit simulation algorithm using these equations that works in  $\mathcal{O}(N^{4})$ time, and in \cref{app:samplecomplexity} we show that for bounded observables (i.e. Pauli strings) the sampling complexity required to get convergence is bounded and independent of system size. We stress that our equations and the corresponding numerical technique is exact, and not the result of any approximations. In particular, it is agnostic to the total amount of entanglement in the system. 

\subsection{Constraints on dynamics}
\label{sec:FPE}

While the above stochastic techniques are more computationally friendly, we can in principle recast \cref{eq:StochasticGamma}, the stochastic equation of motion for our covariance matrix, as a deterministic equation for a probability distribution $p(\Gamma,t)$ on the space of possible covariance matrices. Using this distribution, we could write down the exact density matrix at all times as
\begin{align}
    \hat \rho(t) &= \int p(\Gamma,t) \hat \rho_\Gamma \dd \Gamma, \label{eq:rhoDist}
\end{align}
where $\hat \rho_\Gamma$ is the unique Gaussian state with covariance matrix $\Gamma$. We derive the Fokker-Planck equation governing $p(\Gamma,t)$ in \cref{app:FPE}.  It has a rather unwieldy structure due the point-process noise; in particular, it is not local in the space of covariance matrices. 

The form of \cref{eq:rhoDist} also provides crucial insights into generic features of the class of models we study.  Even though we are able to describe master equations that correspond to non-free fermions (i.e.~where Gaussian initial states can evolve to non-Gaussian states), the dynamics is inherently constrained:  the state at all times can be viewed as a {\it classical mixture} of Gaussian states, described by a classical probability distribution $p(\Gamma, t)$. We note that a similar class of states was recently studied for very different reasons in  \cite{gonzalezgarcia2025}; the motivation here was to simulate a class of interacting fermionic lattice models with strong dephasing.  This mixture-of-Gaussians represents a highly constrained state space. For example, the only pure states in this space are simple Gaussian states.  

We note that this is the same class of states that one would get after time evolving with a classically stochastic quadratic Hamiltonian, which is one way to model quadratic, Hermitian fermionic jump operators. However, despite living in the same state space, we stress that a classically stochastic free Hamiltonian admits significantly simpler dynamics than our particular class of stochastic spin models. A stochastic quadratic Hamiltonian can be modeled by the following equation of motion for the covariance matrix $\Gamma$:
\begin{align}
    \partial_t \Gamma &= -(X_0 \Gamma - \Gamma X_0) \dd t - \sum_i (X_i \Gamma - \Gamma X_i ) \dd W_i.
\end{align}
Here, $\dd W_i$ are independent Wiener increments, $X_0 = 4i  H$ is the static generator of the dynamics, and $X_i = 4i L_i$ are the matrices corresponding to the Hermitian jump operators. This is a very simple set of dynamics, where everything is linear, and the noise is a simple Wiener process independent of the state. This can be contrasted with \cref{eq:StochasticGamma}, where the equation of motion becomes non-linear in $\Gamma$, and the noise itself is a non-linear point process that depends on the state, greatly increasing the complexity of the problem.

Our results also imply that all information on the evolved state (at any time) is encoded in $4N^2$ stochastic degrees of freedom. This should be  contrasted against a  generic state, which requires one to specify $4^N$ degrees of freedom, and alternatively a truly Gaussian state which has $4N^2$ {\it deterministic} degrees of freedom.  Note that a polynomial number of stochastic degrees could still encode an exponential amount of information, if there are an exponential number of non-trivial moments of the distribution function $p(\Gamma,t)$.  However, even in this extreme case, the information is encoded in a more convenient manner.  In particular $(2n)-$point correlation functions are encoded in $n^{th}-$order moments, and are insensitive to higher-order moments. Further, we never need to calculate any given moment larger than order $2N$ (as this is the longest irreducible string of Majorana operators), and so any individual Pauli string is encoded efficiently in the distribution. There are an exponentially large number of these Pauli strings, and so we cannot access \textit{all} information efficiently, but we can access \textit{arbitrary} information efficiently.

Finally, we can also use our equations to derive a useful deterministic but approximate equation of motion for the trajectory-averaged covariance matrix $\BarGamma$.  One simply takes the stochastic average of \cref{eq:GammaEOM}, making a first moment approximation (i.e.~replacing each factor of $\Gamma_{jk}$ with its average on the RHS).  This gives the following approximate equation for the trajectory-averaged covariance matrix (i.e., the covariance matrix of the unconditional state):  
\begin{widetext}
    \begin{align}
    \partial_t \BarGamma_{mn} &=  ( \BarGamma X - X \BarGamma^*)_{mn} +  \sum_{ikl} l^*_{ik} l_{il} (\eta^i_m \eta^i_n -1)(\BarGamma_{km}\BarGamma_{nl} + \BarGamma_{kl}\BarGamma_{mn} - \BarGamma_{kn} \BarGamma_{ml}). \label{eqn:secondCum}
\end{align}
\end{widetext}
While not immediately obvious, this approximation equation matches what one would obtain using a more standard second-cumulant approximation on equations of motion obtained directly from the full fermionic master equation (see \cref{sec:mft} for details).

\section{Numerical Results}
\label{sec:numerics}

In this section we highlight the power of our general method by applying it to understand three paradigmatic examples of open spin chains.  While we focus on 1D examples here, we stress that our technique can also be applied to certain higher-dimensional models (see \cref{app:2d}).  The three examples we study correspond to open spin models where a JW mapping results in non-trivial strings in the dissipators.  In each case, if one simply ignores the strings, one is left with a quadratic fermionic Lindbladian where the physics is easily understood.  The question is then to understand how the inclusion of strings change the physics, i.e.~how is the spin system different from the corresponding free fermion system.  In our examples, the strings will lead to large changes in either the dynamics or the steady state of the system. In the first two examples (AFM order melting and subradiance), we will look at how the addition of the string operators can speed up the approach to the steady state via effective scattering interactions, even if the steady state is trivial vacuum with or without the presence of the string operators. In the final example (an open Ising chain) we will see how the strings can actually affect the steady state observables and effective temperature of the steady state distribution.

\subsection{AFM Order Melting in XX Chains}
\label{subsec:afm}

Our first example is perhaps simplest and seemingly most trivial kind of open spin chain: an XX chain where there is uniform local loss on each site.  This would correspond to a chain of qubits with nearest neighbor couplings, with each qubit having a finite $T_1$ decay time. The system evolves under the master equation given in \cref{eqn:lind}, with

\begin{align}
    \hat H &= -\frac{J}{2} \sum_{i = 1}^{N-1} \left(  \hat \sigma_i^x \hat \sigma_{i + 1}^x + \hat \sigma_i^y \hat \sigma_{i + 1}^y \right), \label{eq:spinAFMHam} \\
    \hat L_i &= \sqrt{\kappa} \hat \sigma_i^- . \label{eq:spinAFMJump}
\end{align}
This system has a unique steady state given by the spin polarized state $|\uparrow \rangle^{\otimes N}$. Furthermore, because the total magnetization is a weak symmetry, we can characterize the time scale for the decay of the total magnetization by observing that if we define $\hat n = \frac{1}{N} \sum_{i = 1}^N \hat \sigma_i^+ \hat \sigma_i^-$, then this decays with a simple exponential to zero:
\begin{align}
    \langle \hat n \rangle(t) &= e^{-\kappa t} \langle \hat n \rangle(0).
\end{align}
This decay profile is identical to what one would find for the dynamics of total density in the corresponding free fermion model, i.e.~a tight-binding chain with singe-particle loss on each site:
\begin{align}
    \hat H &= -J\sum_{i = 1}^{N-1}  \hat c_i^\dagger \hat c_{i + 1} + \hc , \label{eq:fermiAFMHam} \\
    \hat L_i &= \sqrt{\kappa} \hat c_i . \label{eq:fermiAFMJump}
\end{align}
Hence, at first glance, it is tempting to conclude that both the free fermion and spin system (i.e.~fermions with strings in the dissipator) are equally trivial.  Surprisingly, even in this maximally simple model, there are local, extensive observables where the spins and fermions have dramatic differences in their dynamics.  As the average total density is not sensitive to the strings, we instead look at how non-zero Fourier components of the average excitation density decay.  In particular, we focus on N\'eel or anti-ferromagnetic (AFM) order, defined as the $k=\pi$ component of density:
\begin{align}
    \hat A &= \frac{1}{N}\sum_{i = 1}^N (-1)^i (1 - 2\hat c_i^\dagger\hat c_i) = \frac{1}{N}\sum_{i = 1}^N (-1)^i \hat \sigma_i^z.
\end{align}
We consider a scenario where the system is initialized at $t=0$ into the N\'eel state so as to maximize $\langle \hat A \rangle$.  We then study how this order decays to zero.  In the case where the dynamics are given by free fermions, in the thermodynamic limit, we can solve for this exactly. Defining momentum space annihilation operators as
\begin{align}
    \hat{\tilde c}_k &= \frac{1}{\sqrt {N}} \sum_i e^{ijk} \hat c_j,
\end{align}
the AFM order (assuming infinite boundary conditions) is given by
\begin{align}
    A(t) &= \int_{-\pi}^\pi \langle \hat{\tilde c}_k^\dagger \hat{\tilde c}_{k + \pi} \rangle \frac{\dd k}{2 \pi},
\end{align}
where we define $A(t) = \langle \hat A\rangle(t)$. For free fermions, the integrand has an overall decay at rate $\kappa$, and an oscillation set by the energy difference of the momentum states $k$ and $k+\pi$.  This yields
\begin{align}
    A(t) &= \int_{-\pi}^\pi e^{4 i J \cos(k) t - \kappa t} \frac{\dd k}{2 \pi} = 
    e^{-\kappa t} J_0(|4Jt|), \label{eq:fermiBessel}
\end{align}
where $J_0$ is the Bessel function.  The overall exponential decay here just corresponds to the decay of the total particle density.  More interesting is the Bessel function envelope, which is the result of the interplay between the coherent Hamiltonian dynamics and the initial N\'eel ordering.  The simple form here relies on the fact that momentum is conserved (see \cref{app:AFMorder}). At long times, the Bessel function decays like $t^{-1/2}$, giving a power law behavior decay in the scaled correlation $A(t) / e^{-\kappa t}$.  

\begin{figure}[t]
    \centering
    \includegraphics{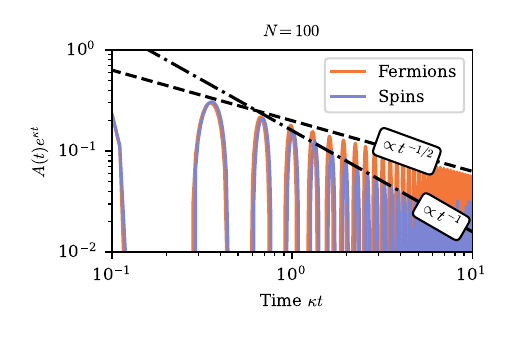}
    \caption{Here we plot the time dynamics of the AFM order parameter melting for either a 1D spin chain [c.f. \cref{eq:spinAFMHam,eq:spinAFMJump}] or free fermions [c.f. \cref{eq:fermiAFMHam,eq:fermiAFMJump}]. The fermion curve corresponds to the analytical solution in the thermodynamic limit given by the Bessel function [c.f. \cref{eq:fermiBessel}], and the spins are found for an $N=100$ spin chain with open boundary conditions averaged over 50,000 trajectories. 
    }
    \label{fig:afm_order}
\end{figure}

We can now turn to the spin model, and again ask how the AFM order decays starting from an initial N\'eel state.  The dynamics now is far more complicated, as the strings in the loss dissipators play a non-trivial role.  In particular, they allow scattering between different fermionic $k$ modes, disrupting the simple dynamics of the purely fermionic model.  Results for a $N=100$ site spin chain using our simulation method are shown in \cref{fig:afm_order}, where we plot the scaled AFM order $A(t) / e^{-\kappa t}$.  While for short times the spin and fermion models show similar behavior (not surprising as few jump have occurred), for larger times $\kappa t \gtrsim 1$ the two systems have very different behavior.  In particular, the decay of the scaled AFM order for the spin model is $\propto 1 / t$, and hence faster than the powerlaw decay  $1/\sqrt{t}$ found in the simple fermion model.  The difference persists for a large range of times, until a longer timescale where boundaries become relevant.  

We can develop intuition for this result and even make approximate quantitative estimates by starting with the Heisenberg/OTOC formulation of the spin-system evolution, where the spins look like the fermions, except every time there is a quantum jump, the sign of one of the bonds in the Hamiltonian flips its sign, leading to a kind of time-dependent disorder.  A very crude way of trying to model this is to consider a fermionic system with static disorder, e.g. a Hamiltonian of the form
\begin{align}
    \hat H &= \sum_i J_i \hat c_i^\dagger \hat c_{i + 1} + \hc,
\end{align}
where the $J_i \in \{J,-J\}$ are random variables.  To get an idea of the decay dynamics, we could fix the disorder level to mimic times in our spin model where $\mathcal{O}(N)$ jumps have occurred. This would be accurate for $\kappa t \gtrsim 1$, which is where the spins and free fermions diverge. In our approximate fermionic model, we can imagine that the disorder will generate some dephasing of the coherence between the $k$ and $k+\pi$ momentum modes that determines the AFM order.  Further, we expect that this dephasing rate will depend on the group velocity of the quasiparticles involved, for the simple reason that for some fixed evolution time,  ``fast moving'' modes will experience more of the spatial disorder than slow moving modes.  This motivates an ansatz for the N\'{e}el order of the form:
\begin{align}
    \langle \hat{\tilde c}_k^\dagger \hat{\tilde c}_{k + \pi} \rangle \sim e^{-4 i J \cos(k) t - \kappa t - 4 \alpha^2 J^2 \sin^2(k) t^2 } ,
\end{align}
where $\alpha$ is some unitless parameter dependent on the correlation length of the noise and $4 J^2 \sin^2(k) t^2 = (\nu_g t)^2$ is the distance squared a wavepacket with momentum $k$ has propagated in time $t$.

Using this form, our disordered fermionic model will have a faster decay of the N\'eel order than the disorder free case, because of the effective dephasing.  We find the decay follows (see \cref{app:AFMorder} for more details):
\begin{align}
    A (t) \sim e^{-\kappa t} \frac{\cos(J t)}{t}.
\end{align}
The faster power-law decay of $1/t$ is in very good agreement with the full numerics for our spin model as shown in \cref{fig:afm_order}, indicating that at least for this model, the effective disorder picture provides good intuition for how spins are different from simple free fermions.  

\subsection{Subradiance}
\label{subsec:subrad}

Our approach is not limited to purely local dissipative processes, but can also describe dissipation correlated across adjacent sites.  This motivates our second example: a 1D XX spin chain where now we have correlated loss on neighboring sites.  This setup can be viewed as a minimal model for studying subradiant decay in an ordered array of 1D emitters, a topic that has received considerable recent interest (see e.g. \cite{Asenjo-Garcia2017,Asenjo-Garcia2017_2,Henriet2019,Albrecht2019,Zhang2019_2,Zhang2020,Zhang2022,Sheremet2023}).  At a basic level, subradiance corresponds to the existence of certain spin-wave modes that have highly suppressed radiative decay due to destructive interference. Such subradiant physics has been studied in the context of a quantum memory for light \cite{Facchinetti2016,Manzoni2018}, as well as having shown suprising fermion-like correlation patterns \cite{Henriet2019}. While this physics is well understood when there is only a single excitation, the physics with multiple excitations is far more complicated and remains an open problem.  Our model and simulation technique provide a setting to study this phenomena with many excitations, in a manner that is numerically exact.    

Our minimal model for subradiance corresponds to a master equation of the form in Eq.~\eqref{eqn:lind} with:
\begin{align}
    \hat H &= - \frac{J}{2}  \sum_{i = 1}^{N-1} \hat \sigma_i^x \hat \sigma_{i + 1}^x + \hat \sigma_i^y \hat \sigma_{i + 1}^y ,\label{eq:spinSubRadHam} \\
    \hat L_i &= \sqrt{\kappa} ( \hat \sigma_i^- + \hat \sigma_{i  +1}^- ). \label{eq:spinSubRadJump}
\end{align}
We note that the 1D subradiant models studied in Refs. \cite{Asenjo-Garcia2017,Asenjo-Garcia2017_2,Henriet2019,Albrecht2019,Zhang2019_2,Zhang2020,Zhang2022,Sheremet2023}, describing ordered arrays of emitters, are derived from a microscopic light-matter interaction, and yield long-range Hamiltonian and dissipative interactions.  In contrast, all our interactions are nearest-neighbor.  There are nonetheless qualitatively many similar features between our models.  All models exhibit a dissipative gap (slowest relaxation rate) that closes in the thermodynamic limit. In our model this gap closing scales as $N^{-2}$ whereas the microscopically-motivated models show a stronger dependence $\propto N^{-3}$.  Further, both in our model and the more microscopic models, there is a set of single-excitation subradiant modes indexed by an integer $j$, with the slow decay rates growing as $\Gamma_j \propto j^2$.

Our correlated decay model will have a trivial steady state solution of the vacuum, but the dynamics are greatly altered by the correlation in the dissipation. The free-fermion equivalent of this model (i.e. the spin model written in terms of JW fermions, but with all string operators dropped) has the form 
\begin{align}
    \hat H &= -J\sum_{i = 1}^{N-1} \hat c_i^\dagger \hat c_{i + 1} + \hc, \label{eq:fermiSubRadHam} \\
    \hat L_i &= \sqrt{\kappa} (\hat c_i - \hat c_{i  +1}) . \label{eq:fermiSubRadJump}
\end{align}
$\Heff$ (which is the same for both models) is diagonalized using momentum modes, resulting in:
\begin{align}
    \Heff &= 2 \sum_k (J\cos(k) - i\kappa \sin^2(k/2)) \hat c_k^\dagger \hat c_k.
\end{align}
Hence, in the thermodynamic $N \rightarrow \infty$ limit, there is a specific momentum mode $k = 0$ that never decays. At the single excitation level, this ``subradiant state'' persists forever with periodic boundary conditions, and for times quadratic in the system size for open boundary conditions. This also leads to anomalous (non-exponential) scaling of local observables in time, due to the fact that the dissipative gap closes at $k = 0$ (i.e., this momentum mode is undamped). 

\begin{figure}[t]
    \centering
    \includegraphics{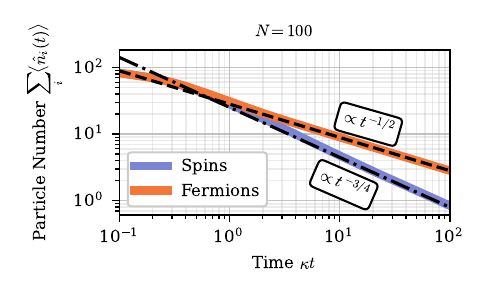}
    \caption{Comparison of the subradiant master equation for spins [c.f. \cref{eq:spinSubRadHam,eq:spinSubRadJump}] and fermions [c.f. \cref{eq:fermiSubRadHam,eq:fermiSubRadJump}] with $J = \kappa = 1$. The free fermion line is given by the analytical expression in \cref{eq:fermiSubRadScaling} and asymptotically scales as $t^{-1/2}$ (dashed black line). The spins are found using the stochastic averaging technique in the Schr\"odinger picture and represents an average of 10,000 trajectories of $N=100$ spins. The local spin number density numerically fits a different power law scaling of $t^{-3/4}$ (dash-dot black line).
    }
    \label{fig:subrad}
\end{figure}

\begin{figure}
    \centering
    \includegraphics[width = \linewidth]{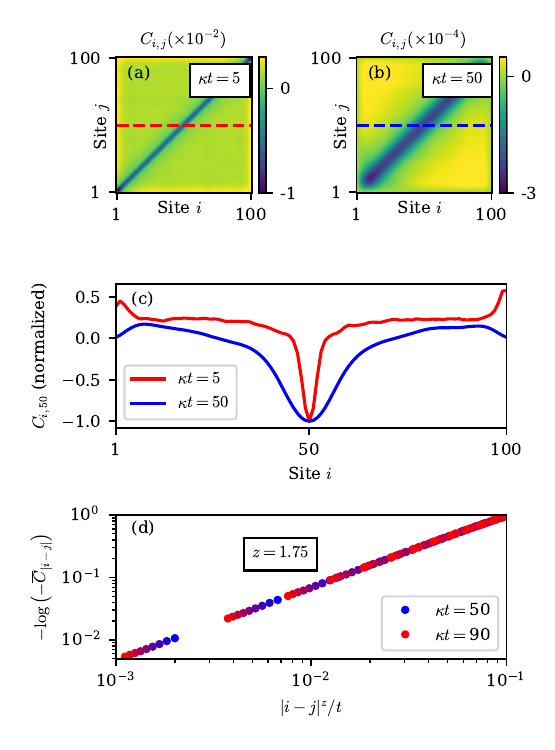}
    \caption{(a) and (b) show the connected density-density correlation functions $C_{i,j}$ as defined in \cref{eqn:concor_main} for the subradiant spin model. The spins are initialized in the all up state, and then the correlation pattern is shown after $\kappa t = 5$ (a) or $\kappa t = 50$ (b). Note the correlations are primarily local, with some boundary effects clearly visible at $\kappa t = 50$. (c) shows the line cuts from (a) and (b) though $i = 50$ in the center of the chain. Each curved is normalized independently such that $C_{50,50}=-1$. As time progresses, the size of the exchange-correlation hole is growing, which can be seen comparing the two line-cuts. (d) shows the logarithm of the windowed averaged correlation function $-\log \left( \overline{C}_{|i-j|}  \right)$ [c.f. \cref{eqn:dd_cor_avg}], plotted for $50 \leq \kappa t \leq 90$ in steps of $\kappa t = 5$. For free fermions, this is simply $|i-j|^2/(2t)$ [see \cref{app:subradiance}], giving a dynamical exponent $z = 2$. We observe scaling collapse for the spins with $z = 1.75$. In all plots $N=100$ lattice sites, and $J = \kappa = 1$, and data is averaged over 10,000 trajectories.}
    \label{fig:dd_cor}
\end{figure}

For concreteness, we imagine starting from the completely-filled state (all qubits in their excited state), and ask how the total number of excitations $n(t)$ relaxes.  For the case of free fermions, there is no scattering between different momentum modes, and so the total density in the thermodynamic limit scales as \cref{app:subradiance}:
\begin{align}
    \langle \hat n \rangle(t) &= \int_{-\pi}^\pi \frac{\dd k}{2 \pi} e^{-2\kappa t\sin^2(k/2)} = e^{-2\kappa t} I_0(2\kappa t), \label{eq:fermiSubRadScaling}
\end{align}
where $I_0$ is the modified Bessel function. Asymptotically, this gives a scaling law of the form $\langle \hat n \rangle(t)|_{t \to \infty}~\sim~(4 \pi \kappa t)^{-1/2}$, see \cref{app:subradiance} for details.  The slow power-law decay directly follows from fact that long wavelength modes with wavevectors $k$ close to $k=0$ have very slow decay rates $\propto k^2$.  Not surprisingly, this result has no dependence on the strength of the exchange coupling $J$.    

Turning now to the spin model (which corresponds to fermions with string operators in the dissipators), the dynamics will in general be more complicated.  One might expect that the above scaling will be modified by the effective interactions encoded by the strings, as now spin waves (i.e.~fermionic $k$ modes) are not conserved but can undergo scattering.  Similar to free fermions, a single excitation prepared in the subradiant mode will not decay (as in this case interactions play no role).  As soon as one has multiple excitations, we expect strong differences.  At a heuristic level, interaction-induced scattering can scatter particles from $k=0$ into finite-$k$ modes, providing a new decay pathway out of the subradiant state.     

In \cref{fig:subrad} we show results for the decay of the average density for a chain of $N=100$ spins, starting from the full excited state and for $J = \kappa$.  We see that the spin model exhibits a faster decay that the free fermion model:  at long times, the decay follows a power law $\langle \hat{n}(t) \rangle \propto (\kappa t)^{-3/4}$, as opposed to the $1/\sqrt{t}$ behavior of free fermions \cite{RyoFootnote}. Here, we again highlight the power of our technique: because there is only loss in the system, the Lindbladian for the free fermions and the spins have an \textit{identical spectrum} \cite{Torres2014}.  The Lindbladian spectrum alone is thus not enough to understand the different decay behavior. Because the dissipative gap closes as a function of $k$, the profile of the Lindbladian eigenmodes is crucial to understanding the long-time power law relaxation in the thermodynamic limit, which we are able to do using only polynomial resources. 

In \cref{app:subradiance}, we show that by running a simulation where we add random telegraph noise to the sign of the couplings in $\Heff$, we can recover the $t^{-3/4}$ power law scaling. This follows from our intuition that the string operators act as conditional gauge transformations, and by adding in random fluctuations to these gauge fields we can qualitatively understand the dynamics of the system, similarly to how adding disorder gave us a qualitative picture of the AFM-order melting in \cref{subsec:afm}. Unlike in the previous section, though, we take the signs of the Hamiltonian couplings to fluctuate in time during the dynamics instead of just inserting static disorder. The reason for this is that, because the subradiant model is gapless, the probability for a jump to occur $\sum_i \langle \hat L_i^\dagger \hat L_i \rangle$ is not decaying exponentially, and so we cannot assume that all jumps happen roughly instantaneously. Instead, there is a non-trivial probability of a jump at any point in the dynamics, giving the telegraph process. 

In order to explore this behavior, it was crucial to be able to simulate very long spin chains, in order to properly distinguish the short time dynamics from the long time dynamics. By using a chain of $N=100$ lattice sites, we are able to show power law scaling over 2 decades during which there is greater than 1 excitation in the system (on average) [see \cref{fig:subrad}]. 

We note that previous work numerically studying subradiance in 1D spin chains did not observe that there was any change in the power law scaling of the mean excitation density with time:  Ref. \cite{Henriet2019} observed that mean particle number for a subradiant spin model decayed in time as $t^{-1/2}$ like the free fermions, whereas we observe $t^{-3/4}$. One possible explanation for this discrepancy is that the model studied in Ref. \cite{Henriet2019} was different than the particular subradiant model we study here, being significantly more collective in both the dissipation and the Hamiltonian interactions. Alternatively, it is also possible that they would also have seen a change in the power-law scaling of the mean density if they had been able to simulate larger system sizes, as we study here, which is closer to the thermodynamic limit. It remains an open question how generic the observed power-law scaling is in other subradiant spin models.

In addition to just having access to the covariance matrix, we also have access to higher weight correlation functions. In previous studies of subradiance, the approximate ``fermionization'' of the dynamics has been explored by studying the connected density-density correlation function
\begin{align}
    C_{i,j} &= \langle \hat n_i \hat n_j \rangle - \langle \hat n_i \rangle \langle  \hat n_j \rangle, \label{eqn:concor_main}
\end{align}
where $\hat n_i = (\hat \sigma_i^z + 1)/2$ is the fermion number density. These density-density correlations were previously discussed in Ref. \cite{Henriet2019}, where they saw the presence of exclusion-like correlation patterns one would expect from a free-fermion state. However, whereas they could look qualitatively at the correlation pattern and see fermion-like behavior, we have a large enough system and long enough times to extract more quantitative data. This correlator can be calculated extremely easily in terms of our fermionized dynamics, because $\hat \sigma_i^z$ is quadratic, and comes with no JW string.

For a system of purely free fermions (neglecting all string operators) the time dynamics of the connected correlation can be calculate explicitly in the long time limit (see \cref{app:subradiance} for details):
\begin{align}
    \frac{C_{i,j}}{\langle \hat n_i \rangle \langle \hat n_j \rangle}(\kappa t \gg |i-j|) &= -e^{-|i-j|^2/2\kappa t}.
\end{align}
Such a pattern of correlation shows a fermionic exchange correlation hole due to Pauli exclusion. Moreover, it has a time dependent length scale $l \sim (\kappa t)^{1/2}$, or in other words there is a dynamical exponent $z = 2$ for the size of the hole, signaling a diffusive process. This can be understood intuitively in the momentum-space picture by noting that in the long time limit, the state is roughly a Gaussian centered around $k = 0$, with a variance in the wavevector scaling like $t^{-1/2}$. Inverting this RMS wavevector in turn gives a length scale, which exactly sets the spatial size of the exchange correlation hole. This would mean that we expect the length scale might go as $l \sim 1/\langle \hat n \rangle$ more generally.

We can now ask what happens in the case of the spin system, which is shown in \cref{fig:dd_cor}. In \cref{fig:dd_cor}(a) and (b) we see a strong correlation-exchange hole appearing at both $\kappa t = 5$ and $\kappa t = 50$, showing the Fermi-like behavior of the 1D subradiant spin chain. Moreover, looking at the line cuts in \cref{fig:dd_cor}(c) we can see that the hole is growing in time, as was expected from the free-fermion system. To be more quantitative on how the exchange correlation hole is growing, let's define the following averaged quantity:
\begin{align}
    \overline{C_{d}} &= 
    \sum_{\substack{s \leq i,j \leq N-s \\ |i-j| = d}} \frac{C_{i,j}}{\langle \hat n_i \rangle \langle \hat n_j \rangle}. \label{eqn:dd_cor_avg}
\end{align}
I.e., we average over a fixed separation $|i-j| = d$, but stay at least $s$ lattice sites away from the edges to mitigate the boundary effects. For the free fermions in the bulk, this should just be $\overline{C_{d}}^\mathrm{free} \sim  -\exp(-d^2/2\kappa t)$. In \cref{fig:dd_cor}(d) we plot the logarithm of this quantity for a system with $N=100$ lattice sites and $s = 20$ lattice sites, and find a scaling collapse with with a superdiffusive dynamical exponent $z = 1.75$, which is different from the free fermion dynamical exponent. Moreover, if we used the naive free fermion picture that the length scale $l \sim 1/\langle \hat n \rangle$, this would imply the length scale should be $l \sim t^{3/4}$, which in turn would imply $z = 4/3 \neq 1.75$ which is incorrect. The role the interactions coming from the strings play is incredibly non-trivial, and affects the dynamical behavior of the exchange-correlation hole. This again stresses the fact that just because the entire spectrum is known, it is still not enough to understand the dynamical phenomenon of the model.

We note that in Ref. \cite{Wang2023b}, they study a class of dissipative free-fermion models that exhibit a tunable dynamical exponent. However none of them are given by $1.75$.

\subsection{Open Ising Model}
\label{subsec:openIsing}

A final system we can consider is an open Ising model with transverse field:
\begin{align}
    \hat H &= \sum_i J \hat \sigma_i^x \hat \sigma_{i + 1}^x + h \hat \sigma_i^z, \\
    \hat L_i &= \sqrt{\kappa} \hat \sigma_i^-.
\end{align}
This model is distinct from the previous two as it has a non-trivial steady state solution. Moreover, if we consider the analogous free fermion model (where we drop the string operators in the jumps):
\begin{align}
    \hat H &= \sum_i J (\hat c_i + \hat c_i^\dagger)(\hat c_{i + 1} - \hat c_{i + 1}^\dagger) + 2h \hat c_i^\dagger \hat c_i, \\
    \hat L_i &= \sqrt{\kappa} \hat c_i,
\end{align} 
then these two master equations will have different steady state solutions. Hence, we can compare directly the impact of the strings on the steady state as opposed to dynamics, as we have done previously.

In order to access steady state quantities using our simulation technique, we initialize the state to an arbitrary Gaussian state (for instance, the vacuum), and then perform time evolution until observables of interest saturate and stop changing in time. 

\begin{figure}
    \centering
    \includegraphics{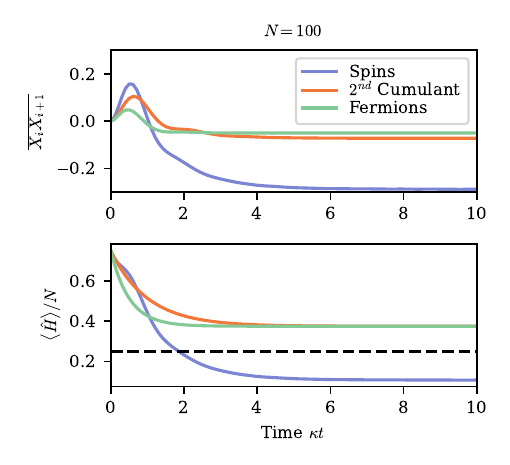}
    \caption{Here, we compare the dynamics of the mean value of the XX interaction $\overline{X_i X_{i + 1}}$ is defined in \cref{eqn:XXCorr} and the average energy is $\langle \hat H \rangle/N = J \langle \hat{C} \rangle + \Delta \langle \hat n \rangle$. We can observe that for all these operators the actual dynamics (blue) differs qualitatively from both the 2nd cumulant expansion (orange) as well as the free fermions (green). The black dashed line corresponds to the infinite temperature energy density $h/2$. Here, we have taken $J = \kappa = 1$, $h = 0.5$ with $N=100$ lattice sites and for the spins we average of 10,000 trajectories.}
    \label{fig:isingFig}
\end{figure}

In previous attempts to study the dissipative Ising model, one technique that has been used is a so-called ``Generalized Gibbs Ensemble'' (GGE) Ansatz \cite{Rigol2007,Ilievski2015,Ali2024}, where it is assumed that the state remains diagonal in the energy eigenbasis of the closed system eigenmodes. A related, less restrictive approximation, is to just make the approximation that the density matrix is Gaussian, and therefore one can use Wick's theorem. As discussed in \cref{sec:FPE}, this actually is equivalent to assuming that the higher moments of the distribution function are all trivial. Such an approximation is well justified in the limit where the dissipation is vanishingly weak \cite{Bouchoule2020,Ali2024}. We also find numerically that it tends to do well when the fixed point is Gaussian, like the previous two examples. However, for this model, when all parameters $J,h,\kappa = \mathcal{O}(1)$, this approximation will qualitatively fail. This can be seen in \cref{fig:isingFig}, where we initialize a completely spin polarized state, and then observe the time dynamics of the average local excitation density, the average nearest neighbor Ising correlation, and the energy density. Interestingly, the mean field theory and the free fermions tend to be quite close together, whereas the spins have a significantly greater value of the nearest neighbor correlation function, and also a significantly smaller average energy density. In fact, if one were to associate the steady state energy density with a temperature, the free fermions and the mean field solution would both have an effective negative temperature, where as the spins relax to a steady state with a positive temperature (energy density smaller than the infinite temperature state). This lower temperature is due almost exclusively to a significantly greater average anti-ferromagnetic correlation than the free fermion solution. If we define the AFM correlation as
\begin{align}
    \overline{X_i X_{i + 1}} &= \frac{1}{N-1} \sum_{i = 1}^{N-1} \langle \hat \sigma_i^x \hat \sigma_{i + 1}^x \rangle , \label{eqn:XXCorr}
\end{align}
then for the spins, this value is roughly 5 times greater than for the free fermions in the steady state, see \cref{fig:isingFig}. The stronger correlations can be understood from the mean field phase diagram (see \cref{app:ising} for more details), which predicts a second order phase transition in the AFM correlation at the parameters we have chosen. Hence, the mean-field value of the correlation would be nearly zero, which is what is seen from the second cumulant expansion as well as the free fermions in \cref{fig:isingFig}. However, as was predicted for much smaller system sizes in \cite{Joshi2013}, the actual spin system doesn't have a sharp transition from ferromagnet to paramagnet, but instead a slow roll off of the nearest neighbor correlations, which is why the actual spin system still has a large amount of correlation.

We also note that the fact that the mean field is qualitatively unable to predict expectation values in this model highlights the power of our numerically exact technique. We believe that the reason mean-field fails in this case is because the steady state of the system can be extremely non-Gaussian. Essentially, we can imagine that the different individual trajectories are able to diffuse out over the space of all Gaussian states. This is in contrast to the previous two models studied, where the steady state was Gaussian. Hence, because the dynamics are forced to begin and end in a Gaussian state, they never ``spread out'' very much in the Gaussian state space, which is why the second cumulant expansion is fairly successful in capturing qualitatively some of the dynamics, as is shown in \cref{app:subradiance}.

In addition to this, the full numerical technique has access to more information than just the quantum steady state, if one takes the unraveling as being physically motivated as opposed to simply a computational tool. For example, if instead of the loss being the result of some simple relaxation process of the spins into the environment, we instead imagine that each spin is being continuously monitored to see whether or not it has decayed, then each trajectory corresponds to an individual measurement record \cite{Breuer2002,Gardiner2004,wiseman2009}. If we now try and ask questions about observables which are not linear in the density matrix, we will get different answers if we look at observables of the averaged state vs averaging over the observable on each trajectory. 

One such example is the entanglement entropy. The density matrix after averaging over trajectories will generically be a mixed state, and so to discuss entanglement entropy requires looking at mixed state entanglement monotones \cite{Bennett1996}. However, each stochastic trajectory remains pure, and so we can meaningfully discuss the entanglement entropy of a given trajectory using more standard means, for example, the Renyi entropies. Of course, before we do so, it is important to ask exactly what we wish to take the entanglement entropy of: we are trying to understand a spin state, but we have used a highly non-local transformation to map it into fermions in order to understand it. \textit{A priori}, the entanglement entropy of the Gaussian fermion state has no relation to the underlying spin model, since the JW is not local. However, as we show in \cref{app:entanglement}, if we take ``good'' bipartitions, the entanglement entropy is actually the same. Here, a good bipartition is one where $A = \{i | 1 \leq i \leq m\}$ and $B = \{i|m < i \leq N\}$ for some chain of length of $N$, where the ordering is defined by the Jordan-Wigner transformation.

\begin{figure}
    \centering
    \includegraphics{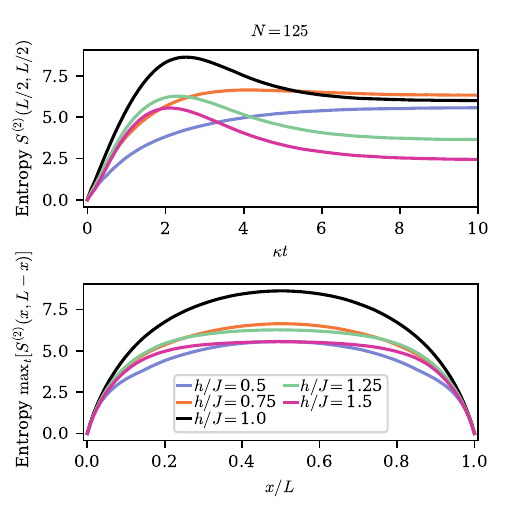}
    \caption{We look at the entanglement entropy on a trajectory level for the open Ising model with $J/\kappa = 10$ and variable $h/J$. We take $N = 125$ lattice sites, and average over 1,000 trajectories, and also quench from an initial product state of all spins down. In the upper plot, we study the bipartite entanglement entropy as a function of time, where when $h/J$ is below the transition, it appears the entanglement is monotonic in time, but it peaks at some early time near the phase transition. In the lower plot, we show Page curves where for each bipartition we look at where the entanglement is maximal as a function of time.}
    \label{fig:isingEntanglement}
\end{figure}

Given this, we can ask how the average bipartite entanglement entropy of each trajectory changes as we tune through the closed system phase transition, and see whether or not the open system is still sensitive to the closed system critical point. This is shown in \cref{fig:isingEntanglement}, where we plot the Renyi-2 entropy for varying values of $h/J$. It appears that when $h/J \lesssim 1$, the entanglement grows monotonically in time before saturating to a steady state value, whereas when $h/J \gtrsim 1$ the entanglement peaks at early times. Moreover, if we look at what the maximal amount of entanglement achieved over all times, the most entanglement seems to be near the closed system phase transition $h/J \sim 1$, and for the system sizes probed the entanglement at this point does not appear to obey an area law. 

\section{ Generalizations }
\label{sec:boson_fermion}

\subsection{Fermions}
The stochastic numerical technique introduced in \cref{sec:new_method} was formulated in order to try and study spin systems that had been Jordan-Wigner transformed into fermions with JW strings. However, there is no reason that it could not be applied directly to a fermionic model that doesn't arise from a spin problem originally (see Refs. \cite{Feng2023,Wang2024,Liu2024} for examples of fermionic master equations where the fact that trajectories remain Gaussian was exploited for simulation). Instead of thinking about string operators, we can instead consider any jump operator of the form of a unitary $\hat U$ multiplied by a Majorana-linear:
\begin{align}
    \hat L_i &= \hat U_i \sum_{j} l_{ij} \hat \gamma_j.
\end{align}
In this case, we have that $\hat L_i^\dagger \hat L_i = \sum_{jk} l^*_{ij} l_{ik} \hat \gamma_j \hat \gamma_k$ is quadratic, and therefore $\Heff$ will still be quadratic in the Majorana operators. 

The only other requirement is that the unitary operator $\hat U_i$ is Gaussian - i.e., it maps quadratic operators to quadratic operators. In the case of the OTOC picture, this means that one can push the unitary onto $\Heff(t)$ that is updated after every jump, and it will remain quadratic. Further, we can also use the state update picture because then the full jump operator remains Gaussian.

If $\hat U_i$ does not preserve Gaussianity, then obviously the state-update picture is not valid, because the state doesn't remain Gaussian over each trajectory. One might be tempted to think that we could instead use the OTOC picture, where now we just have to calculate a much longer OTOC by leaving the unitaries in the expectation value instead of gauging them into $\Heff$. However, unless the unitaries can be written as a single Majorana string (in which case they would be Gaussian anyways), each expectation value for a given observable would require calculating a number of Pfaffians that grows exponentially in the number of jumps, and the number of jumps will be extensize in the system size. Therefore, this reverts us to non-polynomial scaling, and cannot be done.

Thus, the most general type of jump operator that can be simulated efficiently is one where $\hat U_i$ is Gaussian. It is interesting to note that Gaussianity is a gauge dependent choice: an arbitrary unitary rotation the jump operators $\hat L_i \to \sum_j V_{ij} \hat L_j$ (with $V$ a unitary matrix) \textit{does not} affect the unconditional dynamics of the master equation, but it can affect the individual trajectories. For this reason, although the specific unraveling of a master equation is non-unique, there can still be an ``incorrect choice'': a linear combination of Gaussian unitaries need not be a Gaussian unitary. In the spin examples, this can be seen from the fact that taking arbitrary linear combinations of 2-local jump operators need not be 2-local anymore.

\subsection{Bosons}

Let's define the set of bosonic operators $\hat a_i$ that obey the canonical commutation relations $[\hat a_i, \hat a_j^\dagger] = \delta_{ij}$ and $ [\hat a_i, \hat a_j] = [\hat a_i^\dagger, \hat a_j^\dagger]  = 0$. We can similarly consider systems where the jump operator can be written as, for example, $\hat L_i = \hat U_i \hat a_i$, such that $\Heff$ is a quadratic operator and therefore efficiently simulable. In some sense, though, this is a fundamentally different question from the fermions, because linear bosonic operators \textit{do not} preserve Gaussianity. 

Therefore, in order to try and simulate this, it would require working in the OTOC picture. However, we are then left with the problem that, for a Gaussian bosonic state, computing a high-weight observable is equivalent to computing a \textit{hafnian}, and not a \textit{pfaffian} as it is for fermions. While the Pfaffian can be computed efficiently in polynomial time, the Hafnian is \#P-complete, and classically intractable - this underlies the complexity of Gaussian Boson Sampling \cite{Hamilton2017}. Hence, this problem remains intractable. 

Thus, the only problem we could actually simulate for bosons is the case where $\hat L_i  = \hat U_i$ is a unitary operator that preserves the Gaussianity in the state. This, though, is equivalent to stochastically applying a Gaussian unitary channel, which is trivially solvable, as the equations of motion for 2-point correlation functions close on themselves.

\section{Discussion}
\label{sec:disc}

We have introduced a new simulation technique for open quantum systems, with special application to spin chains. Our technique leverages the fact that many models which are interacting at the level of the master equation can be made non-interacting after performing a stochastic unraveling. We show that initially Gaussian states time evolve into states which are non-Gaussian, but can be written as a positive probability distribution over the Gaussian states, which can then be classically sampled from efficiently. This technique opens new avenues in exploring open quantum systems that is not constrained to be either Gaussian or having low-entanglement. Our technique allows access to arbitrary system observables without affecting complexity, even non-local and non-linear ones such as the entanglement entropy.

We have given three concrete examples of 1D spin chains that can now be studied efficiently for large system sizes. First, we observed AFM order parameter melting in an XX chain with local loss, where we saw that the string operators generate effective noise on the Hamiltonian bonds. Next, we studied subradiance in a 1D spin chain with nearest neighbor correlated loss. We observed that, despite not changing the actual spectrum, the effective scattering interactions generated by the string operators can cause faster power-law relaxation to the vacuum than in the case of free fermions. This can be understood by noting that the population in the slowest relaxing modes is no longer conserved, and can be scattered into more quickly relaxing ones. Finally, we have looked at the 1D Ising model with transverse field and single particle loss in the direction of the transverse field. This model has a non-trivial steady state solution which is itself non-Gaussian, and distinct from the steady state one would find if ignoring the string operators. We find that our technique is able to make quantitative predictions about both dynamics as well as the non-Gaussian steady state of this system in regimes where simple approximations (e.g., GGE) fail qualitatively.

We believe that in addition to the simple models outlined above, our technique paves the way to further understanding open 1D spin chains. Our technique not only allows one to numerically probe relevant models, but it also gives insights and intuition as to the effect that string operators can have on the dynamics and steady state of the master equation.

Finally, the technique is not fundamentally limited to only one dimensional systems, and an interesting further direction could be the study of higher dimensional spin models like the Kitaev Honeycomb. Another interesting future direction would be to utilize the non-uniqueness of stochastic unravelings, to see if there are alternative unravelings that are either more optimal numerically \cite{Vovk2022}, or provide different physical insights into the dynamics.

\section*{Acknowledgements}
\label{sec:ack}

We would like to thank Mohammad Maghrebi for useful conversations. This work was supported by the Air Force Office of Scientific Research MURI program under Grant No. FA9550-19-1-0399, the Air Force Office of Scientific Research under Grant No. FA9550-24-1-0354, the Simons Foundation through a Simons Investigator Award (Grant No.~669487), and by the University of Chicago Materials Research Science and Engineering Center, which is funded by the National Science Foundation under Grant No.~DMR-2011854. This work was completed in part with resources provided by the University of Chicago’s Research Computing Center.

\bibliography{ref}

\appendix

\section{Gaussian States Remain Gaussian}
\label{app:gaussiantogaussian}

In this Appendix, we will prove that under each individual trajectory, initially Gaussian states will remain Gaussian. This requires two steps: first, we will demonstrate that time evolution under the effective Hamiltonian preserves Gaussianity, and then we will show that each quantum jump preserves Gaussianity.

We define a Gaussian operator $\hat O_G$ as the exponential of a quadratic fermion operator:
\begin{align}
    \hat O_G &= \exp\left( \sum_{ij} O_{ij} \hat \gamma_i \hat \gamma_j \right). 
\end{align}
A Gaussian state is a Gaussian operator that also is a physical density matrix: i.e., it is positive and has trace one. Gaussian operators form a group under matrix multiplication, which can be seen by observing that commutators of quadratic fermion operators are also quadratic fermion operators, along with application of the Baker-Campbell-Hausdorff formula.

For our general setup, we will take a quadratic Hamiltonian of the form
\begin{align}
    \hat H &= \sum_{ij} H_{ij} \hat \gamma_i \hat \gamma_j, \\
    \hat L_i &= \hat U_i \sum_j l_{ij} \hat \gamma_j,
\end{align}
such that $ \hat U_i^\dagger \hat U_i= \1$ and $\hat U_i$ is a Gaussian unitary operator.

From here, we can observe that the effective Hamiltonian is 
\begin{align}
    \Heff &= \hat H - \frac{i}{2} \sum_i \hat L_i^\dagger \hat L_i
    = \sum_{ij} \left( H_{ij} - \frac{i}{2} L_{ij} \right) \hat \gamma_i \hat \gamma_j,
\end{align}
where $L_{ij} = \sum_k l_{ki}^* l_{kj}$. Thus, the effective Hamiltonian is quadratic, and so the dynamics of the state (up to normalization) is given by
\begin{align}
    \hat \rho_t &\propto e^{-i \Heff t} \hat \rho_0 e^{i \Heff^\dagger t},
\end{align}
which is a product of Gaussian operators and therefore also Gaussian.

The final step is to confirm that the state remains Gaussian following each quantum jump. Since $\hat U_i$ is Gaussian and unitary, then the state $\hat L_i \hat \rho \hat L_i^\dagger$ is Gaussian if and only if $\hat U_i^\dagger \hat L_i \hat \rho \hat L_i^\dagger \hat U_i$. To show that this is a Gaussian operation, we will demonstrate that it is equivalent to appending an auxiliary site, acting with a Gaussian operator, and then projecting the auxiliary onto a Gaussian state. Then, we will show that each of these are themselves a Gaussian operation. 

First, note that adding an auxiliary site in a Gaussian state preserves Gaussianity, as one could always just imagine changing initial conditions and the auxiliary state being non-dynamical, which is equivalent to inserting it in at a later point. Acting with a Gaussian operator is tautologically Gaussian, so it remains to show that we can project onto arbitrary Gaussian states. Let's define $A$ to be our system, $B$ to be the auxiliary site, $\hat \rho_{AB}$ the joint density matrix, and $\1_A \otimes \hat \rho_B$ to be a Gaussian state on $B$. Then we want to show that
\begin{align}
    \hat{\tilde{\rho}}_A &= \tr_B \left( ( \1_A \otimes \hat \rho_B ) \hat \rho_{AB} ( \1_A \otimes \hat \rho_B ) \right),
\end{align}
is also Gaussian. Now, each individual operator within the trace is Gaussian, and so this reduces to showing that a partial trace preserves Gaussianity. Note that
\begin{align}
    \hat \rho_{AB} &\equiv \exp \left[ \left( 
    \begin{array}{c}
        \vec \gamma_A \\
        \vec \gamma_B
    \end{array}
    \right)^T
    \left( 
    \begin{array}{cc}
        \Gamma_{AA} & \Gamma_{AB} \\
        \Gamma_{BA} & \Gamma_{BB}
    \end{array}
    \right)
    \left( 
    \begin{array}{c}
        \vec \gamma_A \\
        \vec \gamma_B
    \end{array}
    \right) \right],
\end{align}
up to normalization. The matrices $\Gamma_{AA}, \Gamma_{BB}$ encode local correlations and must be symmetric to satisfy hermiticity of the state, and the matrices $\Gamma_{AB} = \Gamma_{BA}^T$ encode correlations between system and auxiliary. Explicitly tracing out the state gives us that
\begin{align}
    \hat{\tilde{\rho}}_A &= \tr_B \hat \rho_{AB} \nonumber \\
    &= \mathrm{Pf}(\Gamma_{BB}) \exp \left[ \vec \gamma_A^T (\Gamma_{AA}  - \Gamma_{AB} \Gamma_{BB}^{-1} \Gamma_{BA}  ) \vec \gamma_A\right],
\end{align}
which is by definition a Gaussian state, and so projection and partial trace is a Gaussian operation. 

Now finally, in order to show the action of the quantum jump preserves Gaussianity, we consider adding a Gaussian auxiliary state $\hat \rho_B = |0 \rangle \langle 0|_B$, and then acting with the Gaussian unitary operator 
\begin{align}
    \hat U &= \exp \left( \alpha \hat c_B^\dagger \sum_j l_{ij} \hat \gamma_j - \hc \right).
\end{align}
There are now two relevant cases: if $\sum_j l_{ij} \gamma_j$ is Hermitian, then there is an orthogonal transformation such that we can assume that $\sum_j l_{ij} \hat \gamma_j = \beta \hat \gamma_0$ where $\beta$ is a real normalization. If it is not Hermitian, then we can make a unitary Bogoliubov transformation such that $\sum_j l_{ij} \hat \gamma_j = \beta \hat c_0$. In either case, we can then observe in this new basis that if we take $\alpha = \pi/(2\beta)$, then we can note that
\begin{align}
    \hat{\tilde{\rho}}_A &= \tr_B\left[ (\1 \otimes |1 \rangle \langle 1|_B) \hat U (\hat \rho_A \otimes |0 \rangle \langle 0 |_B ) \hat  U^\dagger (\1 \otimes |1 \rangle \langle 1|_B) \right] \nonumber \\
    &=\left( \sum_j l_{ij} \hat \gamma_j \right) \hat \rho_A \left( \sum_j l_{ij} \hat \gamma_j \right)^\dagger,
\end{align}
as desired.

\section{Set of Simulable 1D Systems}
\label{app:mostgeneral}

Here, we will outline the most general type of 1D spin chains that can be simulated using our technique. We will as before define $N$ to be the total number of lattice sites. The Hamiltonian can be written as
\begin{align}
    \hat H &= \hat H_{\partial} + \hat H_{\mathrm{free}} + \hat H_\mathrm{int}, \\
    \hat H_{\partial} &= \Omega_1 \hat \sigma_1^x + \Omega_2 \hat \sigma_1^y + \Omega_3 \hat \sigma_N^x + \Omega_4 \hat \sigma_N^y, \\
    \hat H_\mathrm{free} &= \sum_i \Delta_i \hat \sigma_i^z + J_{i} \hat \sigma_{i}^+ \hat \sigma_{i + 1}^- + J_{i}^* \hat \sigma_{i}^- \hat \sigma_{i + 1}^+, \\
    \hat H_\mathrm{int} &= \frac{1}{4}\sum_{ij} U_{ij} \left( \hat \sigma_i^z +1 \right) \left( \hat \sigma_j^z + 1 \right),
\end{align}
where $\Omega_i, \Delta_i, U_{ij} \in \R$ and $J_i \in \C$. It is clear that $\hat H_\mathrm{free}$ transforms to free fermions under JW. To understand how to simulate $\hat H_\partial$, we note that we can simply extend the size of the Hilbert space to include two additional lattice sites $i = 0, N +1$ and write a different Hamiltonian \cite{Colpa1979}
\begin{align}
    \hat H'_\partial &= \sigma^x_0(\Omega_1 \hat \sigma_1^x + \Omega_2 \hat \sigma_1^y ) + \hat \sigma_{N + 1}^x (\Omega_3 \hat \sigma_N^x + \Omega_4 \hat \sigma_N^y).
\end{align}
Since $[\hat H'_\partial, \sigma^x_0 ] = [\hat H'_\partial, \sigma^x_{N + 1} ] = 0$ they are constants of motion, and if initialize these states in $|+ \rangle$, then we can formally trace them out and recover $\hat H_\partial$. However, in the new form $\hat H'_\partial$ clearly also transforms to free fermions in an enlarged Hilbert space.

The Hamiltonian $\hat H_\mathrm{int}$ is formally interacting following the JW transformation:
\begin{align}
    \hat H_\mathrm{int} &= \sum_{ij} U_{ij} \hat n_i \hat n_j.
\end{align}
In order to simulate $\hat H_\mathrm{int}$ we use the result from Ref. \cite{gonzalezgarcia2025} that if there is sufficient noise, the quantum channel involving a fermionic density-density interaction (which is what $\hat H_\mathrm{int}$ is) can be written as a map from convex sums of Gaussian states into convex sums of Gaussian states. Hence, this can be simulated in the exact same way as everything else we have simulated by simply adding more stochasticity as we sample Gaussian trajectories. We will not re-derive their result here, but simply quote that the total amount of noise must obey the inequality 
\begin{align}
    \sum_i |U_{ij}| \leq \gamma_j \ \forall j,
\end{align}
where $\gamma_j$ are the set of dephasing rates defined below in \cref{eqn:dephasing}.

Next, we can consider the most general type of jump operators, which can be written as two-local operators that are linear in spin raising/lowering or as dephasing:
\begin{align}
     \hat L_i &= l_{i,1} \hat \sigma_i^- + l_{i,2} \hat \sigma_i^+ + l_{i,3} \hat \sigma_{i + 1}^- + l_{i,4} \hat \sigma_{i + 1}^+, \\
     \hat D_i &= \sqrt{\gamma_i} \hat \sigma_i^z , \label{eqn:dephasing}
\end{align}
where $l_{ij} \in \C$ and $\gamma_j \in \R^+$. We have already discussed at length how one simulates $\hat L_i$. As regards $\hat D_i$, it is simple to note that because the jump operator is Hermitian, we can treat this a stochastic Hamiltonian, i.e. we make the transformation in $\hat H_\mathrm{free}$
\begin{align}
    \Delta_i \to \Delta_i + \sqrt{\gamma_i} \xi_i,
\end{align}
where $\xi_i$ are i.i.d. Gaussian random variables such that $\overline{\xi_i(t) \xi_j(t')} = \delta_{ij} \delta(t - t')$. Hence, $\hat D_i$ can be simulated since $\hat H_{\mathrm{free}}$ can be simulated.

\section{Computational Complexity}
\label{app:complexity}

In this Appendix, we will discuss the computational complexity of both simulation techniques presented in the main text in \cref{sec:new_method}. We will estimate the computational complexity of matrix multiplication of two $N\times N$ matrices as $\mathcal{O}(N^{3})$ using standard algorithms.

\subsection{Heisenberg Picture}

In the Heisenberg picture, we reduce the problem to solving a series of out-of-time-ordered correlators in order to time-evolve a single observable. There are three main steps to the algorithm:
\begin{enumerate}[1.]
    \item Diagonalize $\Heff(t)$.
    \item Time evolve operators within the OTOC using $\Heff(t)$.
    \item Compute the expectation value of an operator at time $t$.
\end{enumerate}
We will define the unitless number $M$ to be the total (expected) number of jumps over a given time trace of time $T$. Implicit within the above three steps is properly identifying when to stochastically insert jump operators, which requires knowing $\langle \hat L_i^\dagger \hat L_i \rangle(t)$ for each jump operator, which we assume to be an extensive number in system size (i.e., there are $\mathcal{O}(N)$ independent jump operators), and so calculating any given time trace requires evolving at least $\mathcal{O}(N)$ operators. The algorithm works as follows: we begin by initializing the number of jumps to be 0, and $\Heff(t) = \Heff(0)$. Then, we repeat the following steps for the desired amount of time:
\begin{enumerate}[1.]
    \item Diagonalize $\Heff(t)$.
    \item Calculate the expectation value of each OTOC for the set of jump operators, using the diagonalized $\Heff$ to time evolve them.
    \item Use these to decide whether or not to do a jump. If so, update $\Heff(t)$, and update the OTOC.
    \item Increment time $\delta t$.
\end{enumerate}
In order to get good convergence, the time increment must be $\mathcal{O}(N^{-1})$, as the probability of having any given jump in a time interval $\delta t$ scales linearly with the number of jumps. Now, we can calculate the computational complexity of the algorithm outlined above. Step 1 is just matrix diagonalization, and therefore is $\mathcal{O}(N^3)$. Step 2 is more costly. If $M$ jumps have occurred, then the OTOC can be written as a string of $2M + 2$ Majorana operators evaluated at different times, assuming the observable $\hat O$ is quadratic. Our OTOC is of the generic form:
\begin{align}
    \mathrm{OTOC} &= \langle \hat \gamma_1(t_1) \dots \hat \gamma_{2M + 2}(t_{2M + 2}) \rangle = \mathrm{Pf} \left( \check M \right), \label{eqn:pf} \\
    \check M_{ij} &\equiv \langle \hat \gamma_i (t_i) \hat \gamma_j (t_j) \rangle  - \delta_{ij}. \label{eqn:M}
\end{align}
To calculate a single entry in the matrix $\check M$ defined in \cref{eqn:M} requires matrix multiplication using the diagonalized form of $\Heff$, making it $\mathcal{O}(N^3)$. Since there are $(2M + 1)^2$ elements of $\check M$, finding the full matrix is therefore $\mathcal{O}(M^2N^3)$. From here, we can calculate the expectation of a single quadratic operator by taking the Pfaffian as in \cref{eqn:pf}. The Pfaffian is $\mathcal{O}(M^3)$, and so the entire procedure is $\mathcal{O}(N^3M^2 + M^3)$. Since this must be done for each jump operator and these are extensive in system size, Step 2 has an overall complexity of $\mathcal{O}(N^4M^2 + NM^3)$

Step 3 requires deciding whether or not to do a jump, which is $\mathcal{O}(N)$ since it requires summing each expectation for each jump operator. Finally, Step 4 is $\mathcal{O}(1)$. All together, we note that in addition, since $\delta t$ muct be $\mathcal{O}(N^{-1})$ the number of steps is extensive with system size, and additionally since $M = \mathcal{O}(N)$ for extensive systems after order 1 time evolution, we find that the total computational complexity is $\mathcal{O}(N^7)$. 

\subsection{Schr\"odinger Picture}

While the previous algorithm gives some intuition on the role the jumps play as dynamical phases, it is computationally quite intractable for even modest system sizes. The state update technique will prove to be significantly better from a computational complexity standpoint. Firstly, we note that $\Heff$ is now time independent. The new algorithm can now be broken into two simple steps:
\begin{enumerate}
    \item Use $\Heff$ to increment the covariance matrix by a single time step $\delta t$.
    \item Use the covariance matrix to decide whether or not to do a jump.
\end{enumerate}
For the first step, we are time evolving according to a matrix equation 
\begin{align}
    \partial_t \Gamma &= \Gamma X - X^* \Gamma  - \frac{1}{2} \Gamma(X - X^*)\Gamma.
\end{align}
In practice, we use a standard fourth-order Runge-Kutta technique. The complexity is just defined by the matrix multiplication, $\mathcal{O}(N^3)$. If we are performing a jump $\hat L_i$, the covariance matrix transforms as
\begin{align}
    \Gamma_{mn} & \mapsto \frac{\sum_{kl} l^*_{ik} l_{il} \tau^i_m \tau^i_n (\Gamma_{km}\Gamma_{nl} + \Gamma_{kl}\Gamma_{mn} - \Gamma_{kn} \Gamma_{ml})}{\sum_{kl} l^*_{ik} l_{il} \Gamma_{kl}}.
\end{align}
Naively, computing this seems to be $\mathcal{O}(N^4)$. However, if we observe that we can define the matrix $G = l^* \Gamma$, then this reduces to 
\begin{align}
    \Gamma_{mn} & \mapsto \tau^i_m \tau^i_n \left( \Gamma_{mn} + \frac{G_{im}G^\dagger_{ni}}{p_i} - \frac{G_{in}G^\dagger_{mi}}{p_i}\right), \\
    p_i &= \sum_{kl} l^*_{ik} l_{il} \Gamma_{kl}.
\end{align}
$p_i$ will already be computed as it is the probability of doing a jump $i$. Calculating $G$ is simply matrix multiplication, which is $\mathcal{O}(N^3)$, and once we have $G$ we only require $\mathcal{O}(N^2)$ steps to update $\Gamma$, giving us a final complexity $\mathcal{O}(N^3)$ for each time step. Since there will again be $\mathcal{O}(N)$ time steps, we arrive at a final complexity of $\mathcal{O}(N^4)$ for a single trajectory. This single trajectory simultaneously calculated the entire covariance matrix. To get access to higher weight observables requires at each time steps taking a Pfaffian. The highest possible weight observable is a string of $\mathcal{O}(N)$ majorana operators, and so the most complex observable requires taking the Pfaffian of an $N \times N$ matrix, which is $\mathcal{O}(N^3)$, and therefore does not affect the computational complexity of the algorithm. Hence, if one is interested in knowing a polynomial number of arbitrary weight Pauli strings, this can be done efficiently.

\section{Efficiently Diagonalizable Liouvillians}
\label{app:classicalDifficulty} 

Here we comment on the results contained in Ref. \cite{Torres2014}, where it is shown that for a certain class of open spin chains, one can find the entire eigensystem for the Lindbladian. Essentially, they show that in systems with pure loss dissipation and Hamiltonians that conserve particle number, then one can block diagonalize the entire Lindbladian $\L$ in the particle number basis such that it is lower triangular. Specifically, let's define the Lindbladian as the sum of two different superoperators:
\begin{align}
    \L &= \hat{\mathcal{H}} + \hat{\mathcal{K}}, \\
    \hat{\mathcal{H}}\hat \rho &= -i\Heff \hat \rho + i \hat \rho \Heff^\dagger, \\
    \hat{\mathcal{K}} \hat \rho &= \sum_i \hat L_i \hat \rho \hat L_i^\dagger.
\end{align}
With this definition, we can see that in the particle number basis, $\hat{\mathcal{H}}$ contains diagonal blocks, and $\hat{\mathcal{K}}$ strictly lowers particle number by 1, so it contains only lower triangular blocks. Hence, the entire spectrum of $\L$ is contained in $\hat{\mathcal{H}}$, which is a quadratic operator of the fermions, and can therefore be diagonalized: the string operators do not change the spectrum of the Lindbladian. However, they can significantly alter the left/right eigenvectors of $\L$. Ref. \cite{Torres2014} also shows that these eigenvectors can be solved for recursively, giving the full eigensystem of the Lindbladian. 

This is an incredibly powerful result; however, as the eigenmodes no longer take a simple form, trying to calculate dynamics of observables is still an exponentially difficult task unless the recursion relations can be solved analytically. Even if one can analytically diagonalize an exponentially large matrix, taking linear combinations of an exponential number of exponentially large vectors still requires exponential resources, even if the ``hard part'' of performing the diagonalization is already done. Such a problem can be seen in, for example, Gaussian boson sampling: one can analytically calculate the full spectral decomposition of the unitary matrix of a Gaussian  boson sampler, but using this to estimate certain observables is still an exponentially difficult task. The power of our technique is that it allows one to efficiently calculate arbitrary observables. 

Additionally, the technique we have presented here is not restricted to pure loss dissipation or particle conserving Hamiltonians, as is seen in the dissipative transverse field Ising model we study in \cref{subsec:openIsing}. Moreover, in \cref{sec:numerics}, we showed that even in a system where the spectra are identical, the effect the strings have on the dynamics can still be very nontrivial.

\section{Estimation Error and Sample Complexity}
\label{app:samplecomplexity}

In order to estimate the sample complexity required to get convergence, it is simple to observe that if we are only interested in calculating strings of Majorana operators $\hat O$, then each operator has a bounded spectrum and so it expectation is bounded above and below: $-1 \leq \langle \hat O \rangle \leq 1$. Hence, application of the Hoeffding inequality implies  that the probability 
\begin{align}
    P(|\mathds{E}_n[O] - O|\geq \epsilon) \leq 2\exp \left( - \frac{n \epsilon^2}{2} \right),
\end{align}
where we have defined $O \equiv \langle \hat O \rangle$ the actual value, and $\mathds{E}_n[O]$ is the arithmetic mean after running $n$ trajectories. Hence, if we desire to estimate $O$ with an error less than $\epsilon$ with a probability of $1 - \delta$, then we require a sample complexity 
\begin{align}
    n \geq \frac{2 \log (2\delta^{-1})}{\epsilon^2}.
\end{align}
Crucially, this estimator does not depend on the size of the system, and therefore does not affect the computational complexity of the algorithm. Moreover, this is a worst case scenario: often we know that each individual sample has a variance significantly smaller than the maximal variance. For example, in systems with pure loss, the total number of particles in the system is tending to zero exponentially quickly. Hence, for example, the variance in the particle number is also tending to zero exponentially quickly in time, and so the number of samples to get a small additive error is also tending to zero.

\section{Fokker-Planck Equation}
\label{app:FPE}

In this Appendix, we will derive a closed form Fokker-Planck Equation (FPE) for the equation of motion of a real, positive, probability distribution over Gaussian states. The basis for this will be the stochastic equation of motion for the covariance matrix derived in \cref{sec:new_method}. However, the exact details of the equation are irrelevant. If we define $\vec X$ to be a vectorized form of the covariance matrix, then we can momentarily forget the physics, and just treat the previously derived equation of motion as a nonlinear SDE driven by a point process. Specifically, it is given by the general form
\begin{align}
\dd X_i &= f_i(X) \dd t + g_{ij} (X) \dd \xi_j, \\
\dd \xi_j &= \left\{ \begin{array}{ccc}
1 & \mathrm{w/ prob} & N_j \dd t \\
0 & \mathrm{w/ prob} &  1 - N_j \dd t
\end{array} \right. .
\end{align}
Let's find the Ito rules for this particular kind of equation of motion. Let's suppose that $\psi(X)$ is an arbitrary function of $X$, and wish to find it's equation of motion. Then we can derive
\begin{align}
\dd \psi &\equiv \lim_{\Delta t \to 0} \psi(X(t + \Delta t)) - \psi(X(t)) \nonumber \\
&=  \lim_{\Delta t \to 0} \frac{ \psi \left( \vec X + \vec f(X) \Delta t + \hat g(X) \Delta \vec \xi \right) - \psi(\vec X) }{\Delta t} \Delta t.
\end{align}
Let's first assume that in a time step $\Delta t$ none of the individual point processes occur. This has probability $1 - \sum_j N_j \Delta t$. With this probability, we set all of the $\xi_j$ to zero, giving
\begin{align}
\lim_{\Delta t \to 0} \frac{ \psi \left( \vec X + \vec f(X) \Delta t  \right) - \psi(\vec X) }{\Delta t} \Delta t = \frac{\partial \psi}{\partial X_i} f_i(X) \Delta t.
\end{align}
Now, suppose that instead, one of the point processes occurs (the odds of multiple is negligible to order $\Delta t$). Then, we have that this term dominates as $\Delta t \to 0$, i.e.
\begin{align}
&\lim_{\Delta t \to 0} \frac{ \psi \left( \vec X + \vec f(X) \Delta t + \hat g(X) \Delta \vec \xi \right) - \psi(\vec X) }{\Delta t} \Delta t \nonumber \\
&= \lim_{\Delta t \to 0} \frac{ \psi \left( X_i + f_i(X) \Delta t + g_{ij}(X) \right) - \psi(\vec X) }{\Delta t} \Delta t \nonumber \\
&=  \psi \left( X_i  + g_{ij}(X) \right) - \psi(\vec X) .
\end{align}
Hence, we can rewrite the equation of motion for $\psi$ as 
\begin{align}
\dd \psi &= \sum_i \frac{\partial \psi}{\partial X_i} f_i(X) \dd t + \sum_j \left[ \psi \left( X_i  +  g_{ij}(X) \right) - \psi(\vec X)  \right] \dd \xi_j.
\end{align}
From here, let's define $\rho(\vec X, t)$ the probability distribution of $\vec X$ at time $t$. By definition, we have that 
\begin{align}
\E[O](t) &\equiv \int  \rho(\vec X, t) O(\vec X) \dd^n \vec X.
\end{align}
\begin{widetext}
Hence, we find that taking an expectation value of the $\dd \psi$ gives us (suppressing sums for brevity such that repeated indices are implicitly summed over)
\begin{align}
\E[\dd \psi](t) &= \int \left[ \rho(\vec X, t) \dd \psi(\vec X) \right] \dd^n \vec X  \nonumber  \\
&= \int \left[  \frac{\partial \psi}{\partial X_i} f_i(X) \rho(\vec X, t) \dd t +  \left[ \psi \left( X_i  +  g_{ij}(X) \right) - \psi(\vec X)  \right] N_j(\vec X) \rho(\vec X, t) \dd t \right] \dd^n \vec X, \\
\implies \int  \dot \rho(\vec X, t) \psi(\vec X) \dd^n \vec X &= \int \left[ \frac{\partial \psi}{\partial X_i} f_i(X) \rho(\vec X, t) +   \left[ \psi \left( X_i  +  g_{ij}(X) \right) - \psi(\vec X)  \right] N_j(\vec X) \rho(\vec X, t) \right]  \dd^n \vec X \nonumber \\
&= \int \left[ \psi(\vec X) \left( \frac{\partial}{\partial X_i}\left[  f_i(X) \rho(\vec X, t) \right] - N_j(\vec X) \rho(\vec X, t) \right)  \right]\dd^n \vec X \nonumber \\
& \ \ \ + \int \left[ \psi \left( X_i  +  g_{ij}(\vec X) \right)  N_j(\vec X) \rho(\vec X, t) \right] \dd^n \vec X.
\end{align}
At this point, let's define the vector valued family of functions $\vec h_j(\vec X) = \vec X + g_{ij}(\vec X) \vec e_i$, where $ \vec e_i$ is the $i^{th}$ unit vector. Let's further assume that these functions are locally invertible. Then, consider the change of variables $\vec Y_j = \vec h_j(\vec X)$. We can define the Jacobian $M_j$ via the set of determinants of matrices $\check M_j$
\begin{align}
M_j &= \det \check M_j, \\
(\check M_j)_{mn} &=  \left( \frac{ \partial (Y_j)_m}{\partial X_n} \right).
\end{align}
Then, we can rewrite the second integral as 
\begin{align}
\sum_j \int \psi \left( X_i  +  g_{ij}(\vec X) \right)  N_j(\vec X) \rho(\vec X, t)  \dd^n \vec X &= \sum_j \int M_j^{-1}  \psi(\vec Y_j) (N_j \circ h_j^{-1})(Y_j) \times \rho(h_j^{-1}(Y_j),t) \dd^n Y_j.
\end{align}
Therefore, we find that
\begin{align}
0 &= \int \psi(\vec X) \left(-\dot \rho(\vec X,t) +  \frac{\partial}{\partial X_i}\left[  f_i(X) \rho(\vec X, t) \right] - N_j(\vec X) \rho(\vec X, t) + M_j^{-1} (N_j \circ h_j^{-1})(X) \times \rho(h_j^{-1}(X),t) \right) \dd^n \vec X.
\end{align}
Because the function $\psi$ is arbitrary, this must hold for all possible functions, and so we get the effective Focker-Plank equation for the distribution
\begin{align}
\dot \rho(\vec X,t) &=  \grad \cdot [\vec f(\vec X) \rho(\vec X,t)] + \sum_j  (N_j \circ h_j^{-1})(X) \times M_j^{-1}(\vec X) \times  \rho(h_j^{-1}(X),t) - N_j(\vec X) \rho(\vec X, t).
\end{align}
Here, the gradient term is the standard drift term in normal FPEs. On the other hand, because we have a point process of discrete jumps instead of the standard Wiener increments, we have a much more complicated ``diffusion'' term, that captures the effect of these stochastic jumps.
\end{widetext}

\section{Second Order Cumulant Expansion}
\label{sec:mft}

The stochastic equation of motion is convenient that it is in closed form, but because it is stochastic it requires averaging. The Fokker-Planck equation has a closed form and is also deterministic, but it is an equation of motion for a $4N^2$ dimensional distribution function, and therefore outside of limiting cases is analytically and numerically somewhat intractable. To make further simplifications, one possibility is making a second order cumulant expansion by assuming that
\begin{align}
    \overline{\Gamma_{ij} \Gamma_{kl}} - \overline{\Gamma_{ij}} \ \overline{\Gamma_{kl}} = 0,
\end{align}
where the overline represents a noise average. If such a simplification were to hold, then one could directly average the stochastic equation of motion for the covariance matrix and get the much simpler closed form, deterministic equation given by
\begin{widetext}
\begin{align}
    \partial_t \Gamma_{mn} &= (\Gamma X - X^* \Gamma  - \frac{1}{2} \Gamma(X - X^*)\Gamma)_{mn} +  \sum_{ikl} l^*_{ik} l_{il} \tau^i_m \tau^i_n (\Gamma_{km}\Gamma_{nl} + \Gamma_{kl}\Gamma_{mn} - \Gamma_{kn} \Gamma_{ml}) - l^*_{ik} l_{il} \Gamma_{kl} \Gamma_{mn} \nonumber \\
    &= (\Gamma X - X^* \Gamma)_{mn} +  \sum_{ikl} l^*_{ik} l_{il} (\tau^i_m \tau^i_n -1)(\Gamma_{km}\Gamma_{nl} + \Gamma_{kl}\Gamma_{mn} - \Gamma_{kn} \Gamma_{ml}).
\end{align}
\end{widetext}
This has the same computational complexity of solving any given individual trajectory, with the added benefit of requiring no averaging.

Quite interestingly, we can observe that this approximation ends up being identical to assuming that Wick's theorem holds at all times $t$ in the state. To see this, let's revert to the distribution function defined in \cref{eq:rhoDist}. We can directly calculate 
\begin{align}
    \langle \hat \gamma_i \hat \gamma_j \hat \gamma_k \hat \gamma_l \rangle &= \int \tr( \hat \rho_\Gamma \hat \gamma_i \hat \gamma_j \hat \gamma_k \hat \gamma_l) p(\Gamma) \dd \Gamma \nonumber \\
    &= \overline{\Gamma_{ij}\Gamma_{kl}} + \overline{\Gamma_{il}\Gamma_{jk}} - \overline{\Gamma_{ik}\Gamma_{jl}},
\end{align}
and therefore, if the second cumulant in the noise vanishes, then Wick's theorem also trivially holds. Numerically, we find that this approximation often works surprisingly well to calculate local expectation values when the fixed point of the dynamics is a Gaussian state. We conjecture that in such cases, because the state must both begin and end in a Gaussian state, the trajectory-averaged state during the time dynamics never deviates too far from being actually Gaussian, and so the approximation does well to capture the dynamics. 

\section{AFM Order Melting}
\label{app:AFMorder}

In this Appendix, we will give more context into the AFM order melting in both the spin system and the free fermions. 

\subsection{Free Fermions}
For free fermions, we can formally make a unitary transformation into plane wave modes:
\begin{align}
    \hat c_k' &= \frac{1}{\sqrt{N}} \sum_{j = 1}^N e^{ijk} \hat c_j,
\end{align}
which allows us to diagonalize simultaneously the Hamiltonian and all the jump operators:
\begin{align}
    \hat H &= 2J \sum_{k = 1}^N \cos(k) (\hat c_k')^\dagger \hat c_k', \\
    \hat L_k' &= \sqrt{\kappa} \hat c_k'.
\end{align}
The initial Ne\'el state has a covariance given by
\begin{align}
    \langle (\hat c_k')^\dagger \hat c_l'\rangle(0) &= \frac{1}{2}(\delta_{kl} + \delta_{k,l + \pi}), \\
    \implies \langle (\hat c_k')^\dagger \hat c_l'\rangle(t) &= \frac{1}{2}(\delta_{kl} + \delta_{k,l + \pi}) e^{2i(\cos k - \cos l)t - \kappa t}.
\end{align}
Fourier transforming back into real space gives us that the AFM order parameter is now given by
\begin{align}
    A(t) &= \frac{1}{N}\sum_i (-1)^i \langle \hat c_i^\dagger \hat c_i\rangle(t) \nonumber \\
    &= e^{-\kappa t} \sum_k e^{-4 i J t \cos(k)} \langle (\hat c_k')^\dagger \hat c_{k + \pi}'\rangle(0) \nonumber \\
    &= \frac{1}{2}e^{-\kappa t} \sum_k e^{-4 i J t \cos(k)}.
\end{align}
If we then take the continuum limit $N\to \infty$, we can replace the sum with an integral to find that, as expected, we get a Bessel function $A(t) = e^{-\kappa t} J_0(|4 J t|)$.

The crucial thing to observe here is that the AFM order is completely determined by the correlations at $k-l = \pi$, and does not depend on the local density.

\subsection{Spins}

When we switch to spins, our intuition tells us that, roughly speaking, we can think of the jumps as a kind of dephasing noise: by randomly flipping signs in the Hamiltonian, we do not ever change the actual particle number, but on average we tend to dephase away correlations. 

For an incredibly simple starting point, let's assume that the only difference between the spins and fermions is that the spins have disorder on the sign of the bond of each Hamiltonian. I.e., let's define the effective Hamiltonian for the spins as
\begin{align}
    \hat H_s &= J \sum_{i = 1}^{N-1} \xi_i \hat c_i^\dagger \hat c_{i +1} + \hc,
\end{align}
where $\xi_i = \{-1,1\}$ is a random variable. Now, because we have taken OBC, we can always gauge these phases away into the operators via
\begin{align}
    \hat{\tilde{c}}_i &= \left( \prod_{j = 0}^{i - 1} \xi_j \right) \hat c_i \equiv \tau_i \hat c_i. \label{eqn:gauge_transform}
\end{align}
Using these new operators, we can rewrite the gauged Hamiltonian as
\begin{align}
    \hat H_s &= J \sum_{i = 1}^{N-1} \hat{\tilde c}_i^\dagger \hat{\tilde{c}}_{i +1} + \hc
\end{align}
Taking infinite boundary conditions, we can diagonalize this simply into plane wave modes. Using this plane wave decomposition, we can then solve for how the underlying degrees of freedom evolve:
\begin{align}
    \langle \hat{\tilde c}_k'^\dagger \hat{\tilde{c}}_l' \rangle &= e^{-\kappa t -2iJ \cos(k) t + 2 i J \cos(l) t}.
\end{align}
Using this expression, along with \cref{eqn:gauge_transform}, we can calculate the disorder averaged expectation value of the plane waves in the ungauged basis:
\begin{widetext}
\begin{align}
\overline{\langle (\hat c'_k)^\dagger \hat c'_{l} \rangle(t)} &= \frac{e^{-\kappa t}}{N^2} \sum_{m,n,q,s} e^{i(km - ln) - i(qm - s n) - 2i J \cos(q) t + 2 i J \cos(s)t } \overline{\tau_m \tau_n } \langle  (\hat c_q')^\dagger \hat c'_{s} \rangle(0) .
\end{align}
\end{widetext}
Now, if $\overline{\tau_m \tau_n} = 1$, we recover exactly the free fermion result, as desired. However, a more realistic noise model would have these phases being incredibly correlated locally: essentially, we are trying to ask about the expected number of jumps that might occur between sites $m$ and $n$ and then averaging over the parity of that number. In the large system size limit, this will approach zero rapidly as $m-n$ becomes large. 

\begin{figure}[t]
    \centering
    \includegraphics{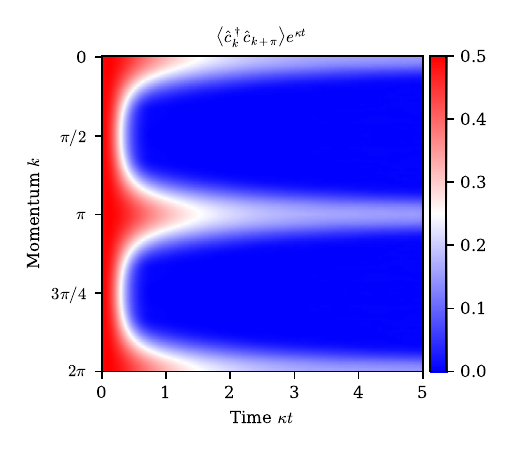}
    \caption{Correlations in momentum space for an $N=100$ site XX spin chain in an initial N\'eel state with local loss on each site. Note that the rate at which correlation decay is dependent on their group velocity.}
    \label{fig:coloredNoise}
\end{figure}

We will posit phenomenologically that $\overline{\tau_m \tau_n} = \exp(-\alpha(m-n)^2)$ for some constant $\alpha$. We know that in order to calculate the AFM order parameter, we simply need to integrate over the correlations at $\pi$. We will again pass to the continuum limit, and observe that this can be rewritten as
\begin{widetext}
\begin{align}
    A(t) &= \frac{1}{2N}e^{-\kappa t} \int_\R \dd x \dd y \int_{-\pi}^\pi \frac{\dd k}{2 \pi} \frac{\dd q}{2 \pi} e^{i(x-y)(k-q)} e^{ - 4iJ \cos(q) t } e^{-\alpha(x - y)^2}.
\end{align}
\end{widetext}
If we rotate the angle into the new coordinates $\bar x = (x +y)$ and $\delta x = (x-y)$, we can observe that $\bar x$ does not show up in the integrand and simply cancels the factor of $N$ in the denominator. We can similarly perform the Gaussian integral in $\delta x$ to arrive at 
\begin{align}
    A(t) &= \frac{1}{2}e^{-\kappa t} \int_{-\pi}^\pi \frac{\dd k}{2 \pi} \frac{\dd q}{2 \pi} e^{-(k-q)^2/(4 \alpha)} e^{ - 4iJ \cos(q) t }.
\end{align}
We can shift the integral over $k \to k - q$ to get that
\begin{align}
    A(t) &= \frac{1}{2}e^{-\kappa t} \int_{-\pi}^\pi \frac{\dd k}{2 \pi} \frac{\dd q}{2 \pi} e^{-k^2/(4 \alpha)} e^{ - 4iJ \cos(k + q) t }.
\end{align}
Now, we make the assumption that $\alpha$ is small enough that we can extend the bounds of integration to infinity with minimal error:
\begin{align}
    A(t) &= \frac{1}{2}e^{-\kappa t} \int_{-\pi}^\pi  \frac{\dd q}{2 \pi} \int_{-\infty}^\infty \dd k e^{-k^2/(4 \alpha)} e^{ - 4iJ \cos(k + q) t }.
\end{align}
This further implies we can Taylor expand around small $k$ to find that
\begin{align}
    A(t) &\sim \frac{1}{2}e^{-\kappa t} \int_{-\pi}^\pi  \frac{\dd q}{2 \pi} \int_{-\infty}^\infty  e^{-\frac{k^2}{4 \alpha}} e^{ - 4iJt[\cos(q) - k \sin(q)] } \dd k \nonumber \\
    &\sim \frac{1}{2}e^{-\kappa t} \int_{-\pi}^\pi  \frac{\dd q}{2 \pi} e^{ - 4iJt \cos(q) - 16 \alpha J^2 \sin^2(q) t^2 }.
\end{align}
This tells us how we expect the off-diagonal elements to dephase: 
\begin{align}
    |\langle (\hat c_q')^\dagger \hat c_{q + \pi}'\rangle(t)|e^{\kappa t} &\sim e^{-16 \alpha J^2 \sin^2(q) t^2}.
\end{align}
Such Gaussian dephasing is common in quasi-static noise processes, and matches extremely well what we see in numerics. Intuitively, it also makes a lot of sense: we can think of the dephasing rate $\alpha J^2 \sin^2(q)$ as the group velocity multiplied by the inverse correlation length of the noise: it tells us how much of the phase randomization a given wave packet samples in time $t$. Wavepackets that move very slowly and have small group velocity take a very long time to notice there is any phase disorder in the lattice, and therefore are much longer lived, and vice-versa for very fast wavepackets. As can be seen in \cref{fig:coloredNoise}, this intuitive picture matches the numerics, where the rate at which the modes are damped out depends on their group velocity. The modes at $k = 0,\pi$ which have zero group velocity are extremely long lived, and the modes at $k = \pi/2,3\pi/2$ with maximal group velocity decay the most rapidly.

We can finally use this expression to derive the correct power law. We wish to calculate
\begin{align}
    A(t) &\propto \int_{-\pi}^\pi  \frac{\dd q}{2 \pi} e^{ - 4iJt \cos(q) - 16 \alpha J^2 \sin^2(q) t^2 }.
\end{align}
At very long times, we can observe that only the values of $q$ near $q=0,\pi$ are relevant. Hence, we can once again approximate this as a Gaussian integral, and extend the bounds to infinity. We also Taylor expand the exponential to get
\begin{align}
    A(t) &\sim \int_{-\infty}^\infty [\cos(4 J t) + 2 J t q^2 \sin(4 J t)] e^{ - 16 \alpha J^2 q^2 t^2 } \dd q \nonumber \\
    &= \frac{\sqrt{\pi}}{4|Jt| \sqrt{\alpha}} \left[ \cos(4 J t) + \frac{\sin(4Jt)}{16 J t \alpha} + \mathcal{O}(t^{-2})\right].
\end{align}
Hence, the leading order scaling is simply $\cos(4 J t)/|Jt|$ up to a prefactor, which agrees extremely well with numerics.

\section{Subradiance}
\label{app:subradiance}

\subsection{Particle Density}

In this Appendix, we will formally derive the long time particle number density for the free fermions, and discuss how this is different for spins. The free fermion master equation is given by
\begin{align}
    \hat H &= J \sum_i \hat c_i^\dagger \hat c_{i + 1} + \hc, \\
    \hat L_i &= \sqrt{\kappa} (\hat c_i - \hat c_{i + 1}).
\end{align}
Using translational invariance, we can Fourier transform this as before to get the new master equation
\begin{align}
    \hat H &= 2J \sum_k \cos(k) (\hat c_k')^\dagger \hat c_k', \\
    \hat L_k &= \sqrt{\kappa}(1 - e^{ik}) \hat c_k' = -2ie^{ik/2}\sin(k/2) \hat c_k'.
\end{align}
If the state is initialized to be completely filled, then each momentum mode will decay independently of all the others at a rate given by
\begin{align}
    \langle (\hat c_k')^\dagger \hat c_k' \rangle(t) &= \exp(-4 \kappa \sin^2(k) t).
\end{align}
From here, it is simple to integrate over all the momentum modes to find the time dynamics of the average particle density:
\begin{align}
    \langle \hat n(t) \rangle &= \int_{-\pi}^\pi \frac{\dd k}{2 \pi} \langle (\hat c_k')^\dagger \hat c_k' \rangle(t)
    = e^{-2\kappa t} I_0(2 \kappa t),
\end{align}
which asymptotically tends to $(4 \pi \kappa t)^{-1/2}$ in the long time limit.

We can think of the long time behavior as effectively integrating over a Gaussian in the long time limit, such that as before we can extend the bounds of the momentum integral to infinity and expand $\sin^2(k/2) \sim k^2/4$ to find the asymptotic behavior. As one might expect, after some time $t$ the only modes that are really contributing have $|k| \lesssim (4 \pi \kappa t)^{-1/2}$, which are still roughly filled. 

If we were to instead try and perform the same calculation with the spins, the crucial difference here is that the different momentum modes are no longer conserved. After some time, the non-interacting master equation is trying to imprint a Gaussian distribution of $k$ modes centered at $k = 0$ with a width $\sigma = (8 \pi \kappa t)^{-1/2}$. However, the interactions allow the different $k$ modes to scatter, which overall tries to make the distribution more uniform. The competition of these two effects - the linear term sharpening the distribution and the non-linear term trying to smoothen it - combine to allow density to leave the system more quickly, which in turn changes the power law.

\begin{figure}[t]
    \centering
    \includegraphics{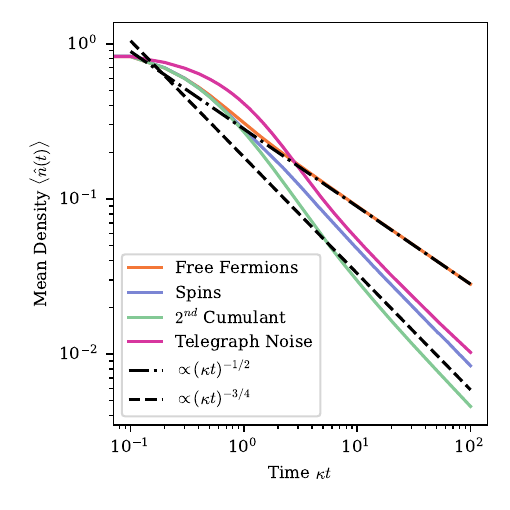}
    \caption{Here, we compare the data in \cref{fig:subrad} for the full spin model (blue) and the free fermion model (orange) to the non-linear time evolution from the second order cumulant expansion method derived in \cref{sec:mft} (green), and the telegraph noise approximation [c.f. \cref{eqn:telegraph_noise}] (magenta). While the actual value of the density at long times is not captured quantitatively, the asymptotic power law is correctly identified to be approximately $(\kappa t)^{-3/4}$ by both the mean field and the telegraph noise.}
    \label{fig:subradMFT}
\end{figure}

We note that this simple scattering picture can be seen at the level of the second cumulant expansion as derived in \cref{sec:mft}. If we define $C'_{kk'} = \langle (\hat c_k')^\dagger \hat c_{k'}'\rangle$ to be the covariance matrix in Fourier space, then the GGE equations of motion are given by
\begin{widetext}
\begin{align}
    \partial_t C'_{qq'} &= -\kappa(\sin^2(q/2) + \sin^2(q'/2)) C'_{qq'} \nonumber \\
    & \ + \frac{\kappa}{N^3} \sum_{k,k',l,l'} \sum_{m,m',n} (\tau_m^n \tau_{m'}^n  -1)e^{i(m(q - l) + m'(l' - q') + n(k' - k)}(1 - e^{i k})(1 - e^{-i k'})(C'_{kk'}C'_{ll'} - C'_{kl'} C'_{lk'}).
\end{align}
\end{widetext}
This is the form of a momentum-conserving 4-point scattering interaction. We can directly time-evolve the second-cumulant equation of motion, which is able to account for the scattering and qualitatively captures the correct power law for the asymptotic decay of the mean particle density, see \cref{fig:subradMFT}. 

Moreover, we can consider an even simpler heuristic model where we think of the action of the spins as causing random classical gauge field fluctuations. I.e., let's consider the following effective Hamiltonian:
\begin{align}
    \Heff &= \sum_i \xi_i \left(J +i\frac{\kappa}{2}\right) \left( \hat c_i^\dagger \hat c_{i + 1} + \hat c_{i +1}^\dagger \hat c_i \right) - i \kappa \hat c_i^\dagger \hat c_i. \label{eqn:telegraph_noise}
\end{align}
Here, $\xi_i \in \{-1,1\}$ is a random telegraph process. Unlike in the AFM-order melting model, this time the dynamics are subradiant, and so we would expect that jumps are continuously happening for times up to $t \sim \mathcal{O}(N^2)$, and so we cannot treat $\xi_i$ as simply static disorder. Instead, we will assume that each $\xi_i$ is a telegraph process where the probability of flipping in a time step $\dd t$ is given by
\begin{align}
    P(\xi_j \to -\xi_j) &= i \frac{\kappa \dd t}{N} \langle \Heff - \Heff^\dagger  \rangle.
\end{align}
Using this model, we are also able to qualitatively capture the dynamics of the spins, as can be seen in \cref{fig:subradMFT}.

\subsection{Density-Density Correlations}

In addition to just studying where the average particle density in the system is, we can also study density-density correlations in the spin chain. We will be interested in the connected correlation function:
\begin{align}
    C_{i,j} &= \langle \hat n_i \hat n_j \rangle - \langle \hat n_i \rangle \langle \hat n_j. \rangle \label{eqn:concor}
\end{align}
For a Gaussian state, this can be calculated using Wick's theorem to observe that $C_{i,j} = -|\langle \hat c_i^\dagger \hat c_j \rangle |^2$. Using this formula, we can first calculate how the connected correlation function should evolve for the free fermion model, which is given by (in the continuum limit, with separation $d = i-j$)
\begin{align}
    C_{i,j}^{\mathrm{free}}(t) &= - \left| \int \frac{\dd k}{2 \pi} e^{ikd - 4 \kappa \sin^2(k/2) t} \right|^2.
\end{align}
This can be computed by expanding the exponential, noting that
\begin{align}
    & \int \frac{\dd k}{2 \pi} e^{-ikd - 4 \kappa \sin^2(k/2) t} \nonumber \\
    &= \sum_{n = 0}^\infty \frac{(-4 \kappa t)^n}{n!} \int \frac{\dd k}{2 \pi} \sin^{2n}(k/2) e^{-i kd} \nonumber \\
    &= \sum_{n = 0}^\infty \frac{(\kappa t)^n}{n!} \int \frac{\dd k}{2 \pi} \left(e^{ik/2} - e^{-ik/2}\right)^{2n} e^{-i kd} \nonumber \\
    &= \sum_{n = 0}^\infty \frac{(\kappa t)^n}{n!} \sum_{j = 0}^{2n} \binom{2n}{j} \int \frac{\dd k}{2 \pi}  e^{ikj/2} e^{-ik(2n - j)/2} e^{-i kd} \nonumber \\
    &= \sum_{n = d}^\infty \frac{(\kappa t)^n}{n!}  \binom{2n}{n + d} = e^{-2 \kappa t} I_d(2 \kappa t),
\end{align}
where $I_d$ is the $d^{th}$ modified Bessel function. This matches on cleanly to the limit $d = 0$ where we recover the value for the local density. 
In the long time limit $\kappa t \gg |i-j|^2$, this can be approximated as
\begin{align}
   C_{i,j}^{\mathrm{free}}(\kappa t \gg |i-j|^2) &\sim -\frac{1}{4 \pi \kappa t} e^{-(i-j)^2/2\kappa t}.
\end{align}
If we then normalize this by the total particle number, the result looks like pure diffusion: 
\begin{align}
    C_{i,j}^{\mathrm{free}}/\langle \hat n_i \rangle \langle \hat n_j \rangle~\to~-e^{-(i-j)^2/2 \kappa t}, \label{eqn:z_ff}
\end{align}
which signifies a dynamical exponent $z = 2$. Instead, the spins appear to observe scaling collapse with a dynamical exponent of $z = 1.75$, signifying superdiffusive behavior.

\section{Open Transverse Field Ising Model}
\label{app:ising}
\begin{figure}
    \centering
    \includegraphics{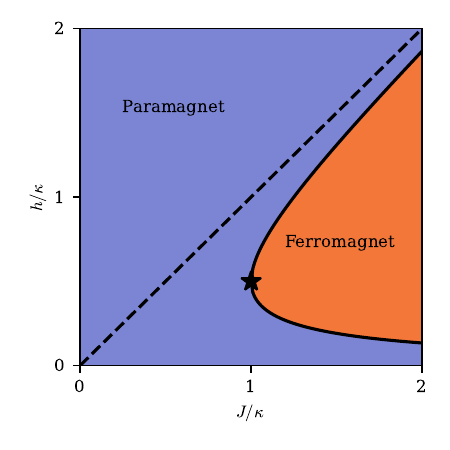}
    \caption{Phase diagram for the mean field dissipative TFIM as defined in \cref{eqn:tfimPD}. The star marks the point $J = \kappa = 2 h$ simulated in the main text \cref{fig:isingFig}. The solid black lines marks the phase boundary for the open system, and the dashed black line is $h = J$ the closed system phase boundary.}
    \label{fig:isingPhaseDiagram}
\end{figure}

In this Appendix, we will derive the mean field phase diagram for the open TFIM. The Hamiltonian and jump operators (as noted in the main text) ares given by
\begin{align}
    \hat H &= \sum_i J \hat \sigma_i^x \hat \sigma_{i + 1}^x + h \hat \sigma_i^z, \\
    \hat L_i &= \sqrt{\kappa} \hat \sigma_i^-.
\end{align}
This gives an equation of motion for the expectation of spin observables:
\begin{align}
    \partial_t \langle \hat \sigma_i^x \rangle &= -h \langle \hat \sigma_i^y \rangle - \frac{\kappa}{2} \langle \hat \sigma_i^x \rangle, \\
    \partial_t \langle \hat \sigma_i^y \rangle &= -J\langle \hat \sigma_i^z \hat \sigma_{i + 1}^x + \hat \sigma_i^z \hat \sigma_{i - 1}^x \rangle +  h \langle \hat \sigma_i^x \rangle - \frac{\kappa}{2} \langle \hat \sigma_i^y \rangle, \\
    \partial_t \langle \hat \sigma_i^z \rangle &= J\langle \hat \sigma_i^y \hat \sigma_{i + 1}^x + \hat \sigma_i^y \hat \sigma_{i - 1}^x \rangle - \kappa \left( \langle \hat \sigma_i^z \rangle - 1 \right).
\end{align}
From here, we make the approximation that expectation values factor $\langle \sigma_i^\alpha \sigma_j^\beta \rangle = \langle \sigma_i^\alpha \rangle \langle \sigma_j^\beta \rangle$, and take the system to be translationally invariant. Then we get the new equations of motion for $\sigma^{x,y,z}$ as
\begin{align}
    \partial_t \sigma^x &= -h \sigma^y - \frac{\kappa}{2} \sigma^x, \\
    \partial_t \sigma^y &= h \sigma^x  - 2J \sigma^x \sigma^z - \frac{\kappa}{2} \sigma^y, \\
    \partial_t \sigma^z &= 2J \sigma^x \sigma^y - \kappa (\sigma^z - 1).
\end{align}
There are two fixed points: the trivial (paramagentic phase when $\sigma^x = \sigma^y = 0, \sigma^z = 1/2)$. We can solve for the other fixed point to find that
\begin{align}
    (\sigma^x)^2 &= \frac{\kappa^2}{8J^2} \left( 4 h J - 4 h^2 - \kappa^2 \right).
\end{align}
This value must be positive, which tells us the phase diagram. Define the scaled parameters $\tilde h = h/\kappa, \tilde J = J/\kappa$. Then we require that
\begin{align}
    \frac{\tilde J - \sqrt{\tilde J^2 - 1}}{2} \leq \tilde h \leq \frac{\tilde J + \sqrt{\tilde J^2 - 1}}{2} . \label{eqn:tfimPD}
\end{align}
In the limit $\kappa \to 0$, then we have that $\tilde J^2 - 1 \sim \tilde J^2$, and we recover the closed system phase transition at $h = 0,J$. Note that we have assumed the sign of $h,J > 0$ here. However, local gauge transformations can move between different sectors, such that the system is ferromagnetic if $h,J$ have the same sign, and antiferromagnetic if they have opposite sign. The phase diagram is shown in \cref{fig:isingPhaseDiagram}.

\begin{figure}
    \centering
    \includegraphics{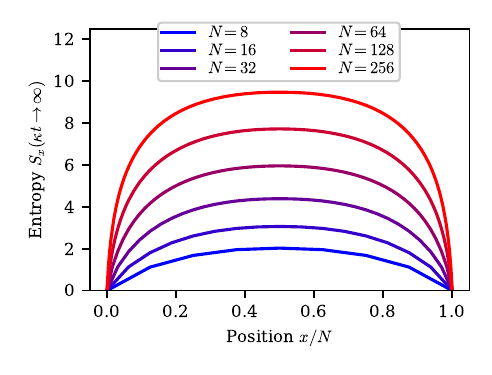}
    \caption{Here, we plot the page curve for the parameter $h = J = \kappa$ as a function of system size from $N=8$ to $N=256$. We are showing the von Neumann entanglement entropy for each partition, averaged over 1000 trajectories.}
    \label{fig:ising256}
\end{figure}

In addition to the mean field phase diagram, we also show data for the unraveled entanglement entropy scaling with system size $N$ in \cref{fig:ising256}.

\section{2D Model}
\label{app:2d}

\begin{figure}
    \centering
    \includegraphics{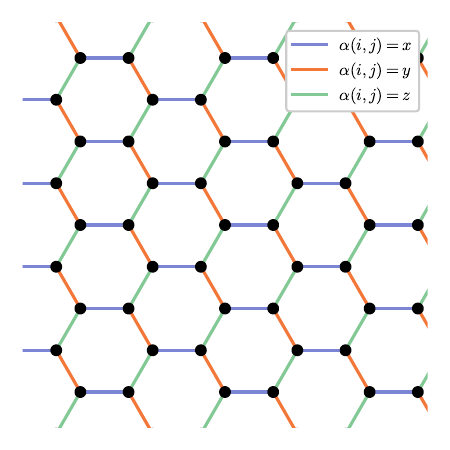}
    \caption{Kitaev Honeycomb model as described in \cref{eqn:honeycomb}}
    \label{fig:honeycomb}
\end{figure}

We have so far exclusively focused on spin chains in a single spatial dimension, as they generically can be transformed into free fermions using the JW transformation. However, there are certain 2D spin models that also have a representation in terms of free fermions, and can thus be extended to open models amenable to our unraveling technique. As an example, consider the Kitaev Honeycomb \cite{Kitaev2006}, a 2D spin model on a hexagonal lattice defined by the Hamiltonian
\begin{align}
    \hat H &= \sum_{\langle i,j \rangle} J_{\alpha(i,j)} \hat \sigma_i^{\alpha(i,j)} \hat \sigma_j^{\alpha(i,j)} , \label{eqn:honeycomb}
\end{align}
where $\langle i,j\rangle$ is a sum over nearest neighbors, and $\alpha(i,j) \in \{x,y,z\}$ depending on the orientation of the bond
(see \cref{fig:honeycomb}). 

This model can be solved by defining on each lattice site 4 Majorana fermions $\hat c, \hat \gamma^\alpha$ (comprising 2 degrees of freedom per lattice site) by
\begin{align}
    \hat \sigma^\alpha_i &= i\hat c_i \hat \gamma_{i}^\alpha.
\end{align}
There is an extra degree of freedom per lattice site, which is fixed by defining the condition:
\begin{align}
    \hat \sigma^x \hat \sigma^y \hat \sigma^z = i \implies \hat c \hat \gamma^x \hat \gamma^y \hat \gamma^z = 1.
\end{align}
Now, we can rewrite the Hamiltonian in the form:
\begin{align}
    \hat H &= \sum_{\langle i,j \rangle} iJ_{\alpha(i,j)} \left( i \hat \gamma_i^{\alpha(i,j)} \hat \gamma_j^{\alpha(i,j)} \right) \hat c_i \hat c_j.
\end{align}
For the moment, this does not look significantly better than it was before: we have a 4-point fermion interaction, which cannot be generically solved. However, if we define the gauge field $\hat A_{ij}$ via
\begin{align}
    \hat A_{ij} &= i \hat \gamma_i^{\alpha(i,j)} \hat \gamma_j^{\alpha(i,j)},
\end{align}
then it is not difficult to check that $[\hat A_{ij}, \hat A_{kl}] = 0$ for all sets of nearest neighbors $\langle i,j\rangle$ and $\langle k,l \rangle$.  Further, these operators all commute with the Hamiltonian. Thus, we can take $\hat A_{i,j} \to A_{ij} = \pm 1$ to be a static background field and write down an effective quadratic theory for the $\hat c$ operators via
\begin{align}
    \hat H &= \sum_{\langle i,j \rangle} iJ_{\alpha(i,j)} A_{ij} \hat c_i \hat c_j.
\end{align}
Hence, if the system is initialized as a Gaussian state in terms of the $\hat c$ operators, then time evolution under this Hamiltonian will keep it Gaussian at all later times, making it efficiently simulable. 

Now, we can consider adding dissipative terms that are local in terms of spin operators that preserve the simulability of the system. For each bond of our lattice, we define the generalized correlated dephasing operators:
\begin{align}
    \hat L_{ij} &= l_{ij}^1 \hat \sigma_i^{\alpha(i,j)} + l_{ij}^2 \hat \sigma_j^{\alpha(i,j)},
\end{align}
where the constants $l_{ij}^{1,2}$ can be complex. If these constants are all real, then this corresponds to a classically stochastic Hamiltonian with local fields driven by white noise. Further, the local fields do not commute with the gauge operators, so this stochastic Hamiltonian would not be solvable with free fermions. Static local fields have previously been studied perturbatively \cite{Kitaev2006}, as they open a gap in the portion of the phase diagram with non-Abelian anyons. However, we show now that the techniques in the main text apply to this model.    

First note that
\begin{align}
    \hat L_{ij}^\dagger \hat L_{ij} &= |l_{ij}^1|^2 + |l_{ij}^2|^2 + 2 \Re \left( l_{ij}^1 (l_{ij}^2)^* \right) \hat \sigma_i^{\alpha(i,j)} \hat \sigma_j^{\alpha(i,j)}, \\
    \implies \Heff &= \hat H -\frac{i}{2} \sum_{\langle i,j \rangle} \hat L_{ij}^\dagger \hat L_{ij} \nonumber \\
    &= \sum_{\langle i,j \rangle} \left[ J_{\alpha(i,j)} - i \Re \left( l_{ij}^1 (l_{ij}^2)^* \right)\right]\hat \sigma_i^{\alpha(i,j)} \hat \sigma_j^{\alpha(i,j)} \nonumber  \\
    &\equiv  \sum_{\langle i,j \rangle} i\tilde J_{\alpha(i,j)} A_{ij} \hat c_i \hat c_j,
\end{align}
and so the effective Hamiltonian remains quadratic with new, complex coupling constants $\tilde J_{\alpha(i,j)} = J_{\alpha(i,j)} - i \Re \left( l_{ij}^1 (l_{ij}^2)^* \right)$. A non-Hermitian Hamiltonian of this form was also studied in Ref.~\cite{Yang2021}. However, they study \textit{only} the non-Hermitian Hamiltonian and completely ignore the quantum jumps. We stress that the no-jump evolution described by $\Heff$ is only relevant to the experimentally-daunting situation where each bond is continuously monitored, and one post-selects on no-jump events.  It is not relevant to our (more general) situation of interest where dissipative effects arise from couplings to unmonitored environments. Moreover, $\Heff$ is trivially solvable as it is completely equivalent to free fermions; the full Lindbladian including the jumps cannot be mapped onto free fermions, which is what we are interested in studying. To do this, the next question to ask is whether or not the system remains Gaussian under a quantum jump. To address this, we write our collapse operators using the previously-introduced Majorana representation:
\begin{align}
     \hat L_{ij} &= i l_{ij}^1 \hat c_i \hat  \gamma_i^{\alpha(i,j)} + i l_{ij}^2 \hat c_j \hat \gamma_j^{\alpha(i,j)}.
\end{align}
Naively, it appears that dissipation generated by such collapse operators will not preserve Gaussianity,
as they are a linear combination of two quadratic operators. 

The crucial observation will be that each $\hat \gamma_i^{\alpha(i,j)}, \hat \gamma_j^{\alpha(i,j)}$ anti-commutes with the local gauge field:
\begin{align}
    \{ \hat \gamma_i^{\alpha(i,j)}, \hat A_{ij} \} &= \{ \hat \gamma_j^{\alpha(i,j)}, \hat A_{ij} \} = 0 .
\end{align}
Hence, when we consider an unraveling, we can think of the action of $\hat \gamma_i^{\alpha(i,j)}, \hat \gamma_i^{\alpha(i,j)}$ as simply locally flipping the sign of the gauge field, sending the \textit{classical variable} $A_{ij} \to -A_{ij}$. To see this more clearly, note that for each bond we can define a Dirac fermion $\hat a_{ij}$:
\begin{align}
    \hat a_{ij} &= \frac{1}{2} \left( \hat \gamma_i^{\alpha(i,j)} + i \hat \gamma_j^{\alpha(i,j)} \right).
\end{align}
We can then explicitly write down the wavefunction for the two different values of the local gauge field as
\begin{align}
    |A_{ij} = -1\rangle &= |0\rangle, \ \ \ |A_{ij} = 1\rangle = \hat a_{ij}^\dagger | 0\rangle,
\end{align}
where $|0\rangle$ is defined as the Dirac fermion vacuum such that $\hat a_{ij} |0\rangle = 0$. Using this explicit basis, we can then compute that
\begin{align}
    \hat \gamma_i^{\alpha(i,j)} |A_{ij} = -1\rangle &= -i \hat \gamma_j^{\alpha(i,j)} |A_{ij} = -1\rangle = |A_{ij} = 1\rangle, \label{eqn:hc_jump} \\
    \hat \gamma_i^{\alpha(i,j)} |A_{ij} = 1\rangle &= i \hat \gamma_j^{\alpha(i,j)} |A_{ij} = 1\rangle = |A_{ij} = -1\rangle. \label{eqn:hc_jump_2}
\end{align}
And so (up to a phase) both $\hat \gamma_i^{\alpha(i,j)}$ and $\hat \gamma_i^{\alpha(i,j)}$ simply flip $\hat A_{ij}$ from one eigenstate to the other eigenstate.

Thus, the action of the jump operator on the state can be understood as moving the state to a different gauge sector. Hence, we can \textit{classically} keep track of the binary variables $A_{ij}$, as the operators $\hat A_{ij}$ will always have a definite value. Every time there is a quantum jump, we first must update the classical bit value for $A_{ij}$, and then we compute the action of the following effective jump operator:
\begin{align}
    \hat{L}'_{ij} &= i l_{ij}^1 \hat c_i - (-1)^{(A_{ij} + 1)/2} l_{ij}^2 \hat c_j,
\end{align}
which are linear and therefore preserve Gaussianity. The sign $(-1)^{(A_{ij} + 1)/2}$ encapsulates the relative phase accumulated in \cref{eqn:hc_jump,eqn:hc_jump_2} where $A_{ij}$ is evaluated pre-jump. Thus, we can imagine that the jump operators are exactly (up to possibly a relative phase) their free-fermion counter parts, along with the fact that every time we do a jump, we update the gauge fields (which also change the Hamiltonian). This is in exact analogy to the 1D case, where we interpret the JW strings as updating the local gauge field at the site of the jump.

\section{Bipartite Entanglement Entropy of the Spin System}
\label{app:entanglement}

Here, we show that the bipartite entanglement entropy for certain bipartitions is independent of whether we calculate it for the spins or the fermions. To see this, let's first observe that the Jordan-Wigner transformation, when viewed as a map from density matrices to density matrices, is unitary. This can be done by picking a basis for both Hilbert spaces, and seeing that JW maps basis vectors to basis vectors. For the spins, this basis is the set of Pauli matrices:
\begin{align}
    \mathcal{B}_s = \{ \hat P_1 \hat P_2 \dots \hat P_N | \hat P_i \in \{ \hat X, \hat Y, \hat Z, \hat I\} \}.
\end{align}
For fermions, we can likewise choose a basis of the set of Majorana operators:
\begin{align}
    \mathcal{B}_f = \{ \gamma_1^{n_1} \hat \gamma_2^{n_2} \dots \hat \gamma_{2N}^{n_{2N}}| n_i \in \{0,1\}\}.
\end{align}
Now, if we ask how a generic operator in the basis $\mathcal{B}_s$ transforms under JW, we can simply note that
\begin{align}
    \hat X_i &\to \left(\prod_{j = 1}^{i - 1} i \hat \gamma_{2j-1} \hat \gamma_{2j} \right) \hat \gamma_{2i-1}, \\
    \hat Y_i &\to \left(\prod_{j = 1}^{i - 1} i \hat \gamma_{2j-1} \hat \gamma_{2j} \right) \hat \gamma_{2i}, \\
    \hat Z_i &\to i \hat \gamma_{2i-1} \hat \gamma_{2i}.
\end{align}
Using the standard anticommutation relations for Majorana's, it is easy to see that any string of Pauli's can be rewritten in the canonical form of an element of $\mathcal{B}_f$ with unit norm. It thus suffices to show the map is surjective to show that this is a unitary map. However, this is also easy by noting that given an arbitrary element of $\mathcal{B}_f$ characterized by a set of binary numbers $\{ n_1, \dots n_{2N}\}$, then we can write the Pauli string $\hat P$ which is it's preimage under JW as
\begin{align}
    \hat P_i &= \left[ \left( \prod_{j = 1}^{i-1} \hat Z_j \right) \hat X_i \right]^{n_{2i-1}} \left[ \left( \prod_{j = 1}^{i-1} \hat Z_j \right) \hat Y_i \right]^{n_{2i}}, \\
    \hat P &= \hat P_1 \hat P_2 \dots \hat P_N.
\end{align}
Hence, the map is a surjective map of basis vectors to basis vectors of equal dimension vector spaces, and is therefore unitary.

Specifically, this means that for any arbitrary density matrix, we can calculate any entanglement monotone from the spectrum of the density matrix, which is unchanged under unitary maps, and therefore is completely unchanged by the Jordan-Wigner transformation. This line of reasoning tells us that entanglement measures on the whole space (i.e. the full state purity) is unaffected by the Jordan-Wigner transform, but we also need to know whether it commutes with taking a partial trace.

As it turns out, for any arbitrary bi-partition of the system, there is no reason the entanglement of the fermions needs to match onto the spin system. However, let's suppose that we take a very special set of bipartitions defined by sets of lattice sites
\begin{align}
    A &= \{ i | 1 \leq i \leq m \}, \\
    B &= \{i | m < i \leq N \}, 
\end{align}
such that $|A| = m, |B| = N-m$. These are perhaps the most natural bipartitions one can imagine taking, such that both subsystems are simply connected using open boundary conditions. Now, let's focus on the $A$ subsystem. In order to understand its entropy, we need to know about the reduced density matrix
\begin{align}
    \hat \rho_A = \tr_B \hat \rho.
\end{align}
Now, we can construct a complete basis for this subspace
\begin{align}
    \mathcal{B}_{s,A} = \{ \hat P_1 \hat P_2 \dots \hat P_m | \hat P_i \in \{ \hat X, \hat Y, \hat Z, \hat I\} \}.
\end{align}
Since this is a basis, specifying the expectation value of these operators completely specifies the state $\hat \rho_A$, and thus the entanglement entropy. However, we can note that under the Jordan-Wigner transformation, we map the set 
\begin{align}
    \mathcal{B}_{s,A} \to \mathcal{B}_{f,A},
\end{align}
where 
\begin{align}
    \mathcal{B}_{f,A} = \{ \gamma_1^{n_1} \hat \gamma_2^{n_2} \dots \hat \gamma_{2m}^{n_{2m}}| n_i \in \{0,1\}\},
\end{align}
and so again on this restricted subspace, we are simply unitarily mapping basis vectors to basis vectors, and the full information of the spin reduced density matrix is completely contained in the fermionic reduced density matrix on the same sublattice. Thus, the two states have the same entanglement entropy, and we can calculate the fermionic one (which is much simpler since the state is Gaussian) in order to learn what the entanglement in the spin system is.

Note that this argument breaks down when $A$ is not a simply connected subset, since the JW string would overlap with the complimentary subspace $B$, and so the total amount of information of the reduced spin system would be encoded in fermionic operators with support in both sublattices.

\end{document}